\documentclass{aa}

\usepackage{graphicx}
\usepackage{natbib}
\usepackage{siunitx}
\usepackage{nicefrac}
\usepackage{multirow}
\usepackage[draft]{hyperref}
\bibpunct{(}{)}{;}{a}{}{,}

\begin{document}

    \title{Influence of metallicity on the near-surface effect on oscillation frequencies}

    \author{L. Manchon\inst{1,2}\thanks{\emph{Present address:}
        Institut d'Astrophysique Spatiale, Université Paris Sud, Orsay, France}
        \and K. Belkacem\inst{2}
        \and R. Samadi\inst{2}
        \and T. Sonoi \inst{2, 3}
        \and J. P. C. Marques\inst{1}
        \and H.-G.~Ludwig\inst{4,5}
        \and E. Caffau\inst{5}}

    \institute{Institut d'Astrophysique Spatiale, Université Paris-Sud, Orsay, France \\ 
    \texttt{email: louis.manchon@ias.u-psud.fr}
    \and LESIA, Observatoire de Paris, PSL Research University, CNRS, Université Pierre et Marie Curie, Université Denis Diderot, 92195 Meudon, France
    \and Astronomical Institute, Tohoku University, 6-3 Aramaki Aza-Aoba, Aoba-ku Sendai, 980-8578, Japan
    \and Zentrum für Astronomie der Universität Heidelberg, Landessternwarte, Königstuhl 12, D-69117 Heidelberg, Germany
    \and GEPI, Observatoire de Paris, PSL University, CNRS, 5 Place Jules Janssen, 92190 Meudon, France}

    \date{Received XXX / Accepted XXX}

    \abstract   
                {The CoRoT and \textit{Kepler} missions have provided high-quality measurements of the frequency spectra of solar-like pulsators, enabling us to probe stellar interiors with a very high degree of accuracy by comparing the observed and modelled frequencies. However, the frequencies computed with 1D models suffer from systematic errors related to the poor modelling of the uppermost layers of stars. These biases are what is commonly named the near-surface effect. The dominant effect is thought to be related to the turbulent pressure that modifies the hydrostatic equilibrium and thus the frequencies. This has already been investigated using grids of 3D hydrodynamical simulations, which also were used to constrain the parameters of the empirical correction models. However, the effect of metallicity has not been considered so far.}
                {We aim to study the impact of metallicity on the surface effect, investigating its influence across the Hertzsprung–Russell diagram, and providing a method for accounting for it when using the empirical correction models.}
                {We computed a grid of patched 1D stellar models with the stellar evolution code CESTAM in which poorly modelled surface layers have been  replaced by averaged stratification computed with the 3D hydrodynamical code CO$^5$BOLD. It allowed us to investigate the dependence of both the surface effect and the empirical correction functions on the metallicity.}
                {We found that metallicity has a strong impact on the surface effect: keeping $T_{\rm eff}$ and $\log g$ constant, the frequency residuals can vary by up to a factor of two (for instance from $[\rm Fe/H] = +0.0$ to $[\rm Fe/H] = +0.5$). Therefore, the influence of metallicity cannot be neglected. We found that the correct way of accounting for it is to consider the surface Rosseland mean opacity. It allowed us to give a physically grounded justification as well as a scaling relation for the frequency differences at $\nu_{\rm max}$ as a function of $T_{\rm eff}$, $\log g$ and $\kappa$. Finally, we provide prescriptions for the fitting parameters of the most commonly used correction functions.}
                {We show that the impact of metallicity through the Rosseland mean opacity must be taken into account when studying  and correcting the surface effect.}
    
    \maketitle

\section{Introduction}

The space-borne missions CoRoT \citep{Baglin2006,Michel2008,Auvergne2009} and \textit{Kepler} \citep{Borucki2010} have provided a rich harvest of high-quality seismic data for solar-like pulsators. This has allowed a leap forward in our understanding and modelling of low-mass stars \citep[see the reviews by][]{ChaplinMiglio2013,HekkerJCD2017}. 
However, for the last three decades \citep[e.g.][]{Dziembowski1988} it has been known that the comparison between modelled and observed acoustic-mode frequencies suffer from systematic discrepancies. This bias is called the surface effect and has been widely studied in the solar case  \citep{Rosenthal1995,Christensen-Dalsgaard1997,Rosenthal1999}. They are attributed to our deficient modelling of the uppermost layers of stars with a convective envelope. Indeed, 1D stellar models hardly take into account the complexity of these layers that are subject to highly turbulent flows as well as a complex transition between a convective to a radiative-dominated energy flux transport \citep[e.g.][]{Kupka2017}. 

More generally, these frequency residuals prevent a direct comparison between modelled and observed frequencies. Frequency combinations are commonly used to circumvent this problem \citep[e.g.][]{Roxburgh2003}, but still, an accurate determination of frequencies is highly desirable to take advantage of the full potential of asteroseismology. To reach this goal, a handful of empirical prescriptions with adjustable free parameters have been proposed \citep{Kjeldsen2008,Ball2014,Sonoi2015,Ball2017} and allow one to apply a posteriori corrections to the modelled frequencies. Such an approach is now widely used \citep[e.g.][]{Lebreton2014,Aguirre2017} and has proven to be quite efficient in inferring a stellar model that fits the observed frequencies. However, it suffers from some fundamental drawbacks. The choice of the parameters is not physically motivated. Consequently, there is no guarantee that this optimal model is unique and accurate (i.e. that it properly reproduces the real physical structure of the observed star).

Another complementary approach then consists of investigating the physical nature of the surface effect. This motivated a number of studies to unveil and constrain the physical ingredients responsible for these biases. More precisely, surface effect has been shown to be the result of two distinct effects \citep[e.g.][]{Houdek2017}: \textit{structural} effects coming mainly from turbulent pressure in the hydrostatic equation which is usually absent in 1D stellar evolution codes, and \textit{modal} effects gathering modifications of the eigenmodes, mostly due to non-adiabaticity \citep[e.g.][]{Balmforth1992a,Houdek2017} as well as the perturbation of turbulent pressure induced by the oscillations \citep{Sonoi2017}. Other related processes were also invoked, such as convective backwarming \citep{Trampedach2017} or magnetic activity \citep{Piau2014,Magic2016}. Nonetheless, as demonstrated by the early work by \citet{Rosenthal1999} on the Sun using a 3D hydrodynamical simulation, the dominant physical ingredient is thought to be the turbulent pressure that modifies the hydrostatic equilibrium and subsequently introduces an elevation of the star surface. Then, the acoustic cavity is modified and therefore the frequencies are as well.

Based on a grid of 3D numerical simulations, this method was used by \citet{Sonoi2015,Ball2016,Trampedach2017} who investigated the surface effect variations across the Hertzsprung-Russell diagram. These works clearly demonstrated that surface effects sharply depend on effective temperature and surface gravity of star. In addition, \citet{Sonoi2015} presented a way to provide parameters for the empirical surface corrections by fitting them against a physically motivated scaling relation derived by \citet{Samadi2013}. However, all these works considered solar metallicity models while the distribution of metallicity for observed solar-like pulsators is quite large \citep[see e.g.][]{Pinsonneault2014}. Our goal is thus to study the influence of metallicity on the surface effects and propose a method to account for it.  

The article is organized as follows: in Sect. \ref{section:s2} we describe the method of model patching , which is constructed by replacing the upper layers of a 1D model by horizontally averaged stratification of a 3D model atmosphere, and our set of models. Then in Sect. 3 we show that metallicity has a strong impact on the frequency residuals and therefore its influence cannot be ignored. We also study the variation of the frequency differences with effective temperature, surface gravity and opacity and give a theoretical justification. Finally, in Sect. 4 we provide constraints on the various parameters usually used in the empirical surface effect function across the $T_{\rm eff} - \log g - \log \kappa$ space.

\section{Model-patching method}
\label{section:s2}
In this section we explain the method we used to patch our models and describe our final set of models.
 
\subsection{Grid of 3D models} 
\label{sub:patched_models_calculation}

\begin{table*}
    \caption{Characteristics of the 3D hydrodynamical models and of the UPM and PM. The final three letters of the model labels correspond to $[\rm Fe/H]$: (m00, m05, m10, p05) refer to $[\rm Fe / H] = (0.0, -0.5, -1.0, +0.5)$, respectively. $T_{\rm b}$ is the mean temperature at the bottom of the 3D model and $\nu_{\rm max}$ is the frequency with the largest amplitude in the oscillation power spectrum ($\nu_{\rm max} = 3050.0 (M/M_\odot)(R_\odot^2/R^2)(5777/T_{\rm eff})^{1/2}$, \citet{Kjeldsen1995}) and $M$ is the mass of the PM which differs by a fraction $\lesssim 10^{-7}$ from the one of UPM. The initial helium and metal abundances are close to the ones at the surface. We recall that the metal abundance is different from the iron abundance $[\rm Fe/H]$ imposed in our models. The evolutionary stages  PMS, MS, and SG stand for pre-main-sequence, main-sequence and sub-giant.}  
    \label{table:3d_param}     
    \centering
    \begin{tabular}{l | c c c c | c c c c c c c c}       
        \hline\hline                   
Model & $T_{\rm eff}$ & $\log g$ & $T_{\rm b}$         & $\nu_{\rm max}$ & $M$         & Age            & $Y_{\rm init}$ & $Z_{\rm init}$ & $\alpha_{\rm MLT}$ & $R_{\rm PM}$ & $\Delta r/R_{\rm PM}$ & Stage \\
      & K             &          & $(\times 10^{4})$ K & $\mu$Hz         & $[M_\odot]$ & $[\mbox{Myr}]$ & $\%$           & $\%$           &                   & $[R_\odot]$  & $(\times 10^{-3})$    &  \\ 
        \hline
Cp05 & $6443$ & $4$     & $3.20$  & $1057$  & $1.91$  &$1341$   & $24.2$ & $4.14$   & $1.61$ & $2.29$  & $1.3$    & MS\\
Fp05 & $6188$ & $4$     & $2.71$  & $1078$  & $1.81$  &$1820$   & $24.2$ & $4.14$   & $1.68$ & $2.23$  & $0.89$   & MS\\
Jp05 & $5877$ & $4$     & $2.17$  & $1103$  & $1.53$  &$4129$   & $24.2$ & $4.16$   & $1.74$ & $2.05$  & $0.67$   & SG\\
Lp05 & $5431$ & $4$     & $2.03$  & $1161$  & $1.32$  &$8361$   & $24.2$ & $4.14$   & $1.74$ & $1.89$  & $0.45$   & SG\\
Op05 & $4933$ & $4$     & $1.98$  & $1205$  & $2.12$  &$3.808$  & $24.2$ & $4.14$   & $1.75$ & $2.41$  & $0.16$   & PMS\\
 \hline
Am00 & $5776$ & $4.44$  & $1.55$  & $3063$  & $1.02$  &$4628$   & $24.9$ & $1.35$   & $1.65$ & $1.01$  & $0.21$   & MS\\
Bm00 & $6730$ & $4.25$  & $7.89$  & $1839$  & $1.38$  &$1306$   & $24.9$ & $1.35$   & $1.68$ & $1.46$  & $1.4$    & MS\\
Cm00 & $6490$ & $4$     & $2.79$  & $1053$  & $1.48$  &$2397$   & $24.9$ & $1.35$   & $1.65$ & $2.01$  & $1.3$    & MS\\
Dm00 & $6434$ & $4.25$  & $2.76$  & $1881$  & $1.27$  &$2562$   & $24.9$ & $1.35$   & $1.67$ & $1.4$   & $0.77$   & MS\\
Fm00 & $6231$ & $4$     & $2.30$  & $1074$  & $1.28$  &$4352$   & $24.9$ & $1.35$   & $1.69$ & $1.87$  & $0.96$   & SG\\
Gm00 & $6182$ & $3.5$   & $3.63$  & $341.7$ & $1.93$  &$1158$   & $24.9$ & $1.35$   & $1.7$  & $4.09$  & $2.4$    & SG\\
Hm00 & $6101$ & $4.25$  & $2.51$  & $1923$  & $1.13$  &$5413$   & $24.9$ & $1.35$   & $1.67$ & $1.32$  & $0.53$   & SG\\
Im00 & $5861$ & $4.5$   & $2.31$  & $3479$  & $1.09$  &$34.8$   & $24.9$ & $1.35$   & $1.66$ & $0.976$ & $0.2$    & PMS\\
Jm00 & $5936$ & $3.99$  & $2.18$  & $1077$  & $1.16$  &$6753$   & $24.9$ & $1.35$   & $1.74$ & $1.8$   & $0.51$   & SG\\
Km00 & $5886$ & $3.5$   & $2.18$  & $348.5$ & $1.88$  &$1249$   & $24.9$ & $1.35$   & $1.68$ & $4.05$  & $1.5$    & SG\\
Mm00 & $5436$ & $3.5$   & $2.03$  & $364.4$ & $2.21$  &$1017$   & $24.9$ & $1.35$   & $1.71$ & $4.37$  & $0.83$   & SG\\
 \hline
Em05 & $6227$ & $4.5$   & $2.22$  & $3395$  & $1.02$  &$644.2$  & $25.1$ & $0.597$  & $1.74$ & $0.938$  & $0.24$   & MS\\
Fm05 & $6260$ & $4$     & $2.26$  & $1079$  & $1.06$  &$6520$   & $25.1$ & $0.597$  & $1.79$ & $1.7$    & $0.85$   & SG\\
Gm05 & $6149$ & $3.5$   & $3.27$  & $338.2$ & $1.66$  &$1389$   & $25.1$ & $0.597$  & $1.76$ & $3.82$   & $2$      & SG\\
Im05 & $5897$ & $4.5$   & $2.25$  & $3502$  & $0.902$ &$4290$   & $25.1$ & $0.597$  & $1.62$ & $0.882$  & $0.17$   & MS\\
Jm05 & $5919$ & $4$     & $2.19$  & $1091$  & $0.949$ &$10~130$ & $25.1$ & $0.597$  & $1.76$ & $1.62$   & $0.52$   & SG\\
Km05 & $5764$ & $3.51$  & $2.13$  & $363.2$ & $1.59$  &$1623$   & $25.1$ & $0.597$  & $1.74$ & $3.66$   & $1.5$    & SG\\
Mm05 & $5463$ & $3.5$   & $2.01$  & $359.4$ & $1.63$  &$1483$   & $25.1$ & $0.597$  & $1.74$ & $3.78$   & $0.7$    & SG\\
 \hline
Bm10 & $6743$ & $4.24$  & $7.60$  & $1801$  & $0.776$ &$12~030$ & $25.3$ & $0.0205$ & $1.84$ & $1.1$     & $1.2$    & SG\\
Am10 & $5771$ & $4.44$  & $1.53$  & $3065$  & $0.768$ &$13.47$  & $25.3$ & $0.0205$ & $1.26$ & $0.875$   & $0.32$   & PMS\\
Cm10 & $6503$ & $4$     & $2.63$  & $1052$  & $0.765$ &$13~770$ & $25.3$ & $0.0205$ & $1.75$ & $1.45$    & $1.4$    & SG\\
Fm10 & $6242$ & $4$     & $2.25$  & $1073$  & $1.17$  &$3.639$  & $25.3$ & $0.0205$ & $1.67$ & $1.79$    & $0.8$    & PMS\\
Gm10 & $6213$ & $3.5$   & $3.14$  & $341$   & $1.03$  &$5093$   & $25.3$ & $0.0275$ & $1.7$  & $2.99$    & $3$      &  SG\\
Km10 & $5894$ & $3.5$   & $2.15$  & $349.7$ & $0.906$ &$7858$   & $25.3$ & $0.0205$ & $1.62$ & $2.8$     & $1.9$    & SG\\
        \hline
    \end{tabular}
\end{table*}

We used a grid of 3D hydrodynamical models from the CIFIST grid of stellar atmosphere including the superadiabatic region to the shallowest layers of the photosphere, computed using the CO$^5$BOLD code \citep[see][for details]{Ludwig2009,Freytag2012}.  The chemical mixture is based on the solar abundances of \citet{Grevesse1998} apart from the CNO elements which follow \citet{Asplund2005}. We considered 29 models with effective temperature ($T_{\rm eff}$) ranging from $4\,500$ K to $6\,800$ K, a surface gravity ($\log g$) ranging from $3.5$ to $4.5$, and a metallicity $[\rm Fe/H]=\{-1.0, -0.5, +0.0, +0.5\}$. We note that $[\rm Fe/H]$ refers to the logarithmic iron abundance, which in our simulations is different from the logarithmic metallicity abundance $\rm [M/H]$. Thus, models with same $\rm [Fe/H]$ do not necessarily have the same $\rm [M/H]$.

\begin{figure}
    \includegraphics[width=\hsize]{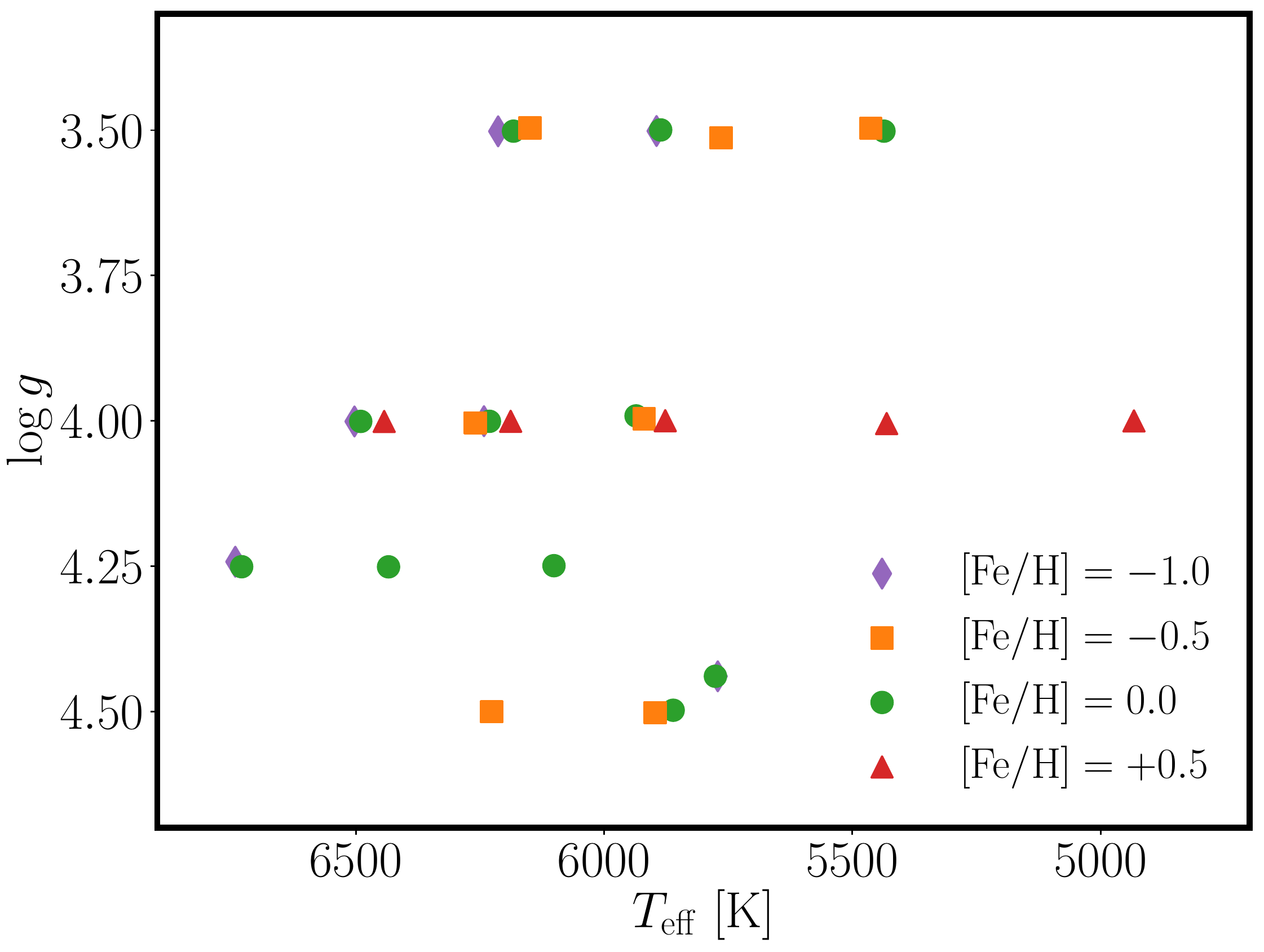}
    \caption{Patched models represented in Kiel diagram (see Sect. \ref{sub:patched_models_calculation}). The metallicities are shape- and colour-coded and are $[\rm Fe/H]=\{-1.0, -0.5, +0.0, +0.5\}$.}
    \label{fig:patched_models}
\end{figure}

Table \ref{table:3d_param} summarizes the global parameters of the 3D models. The range of metallicities we considered corresponds to the metallicities of observed solar-like pulsators \citep{Anders2017,Serenelli2017}. Table \ref{table:3d_param} exhibits small groups of models (labelled with same first letter) with very similar $T_{\rm eff}$ and $\log g$. Those groups\, for instance in Fig. \ref{fig:patched_models} at $\log g =4.0$, help us to investigate the influence of metallicity on the surface effect by keeping other global parameters fixed. However, we pointed out that, whereas within a group the dispersion in $\log g$ is rather small (of the order of $0.1\%$), the dispersion in $T_{\rm eff}$ is much higher (of the order of $1\%$). Indeed, surface gravity is an input parameter of the hydrodynamical simulations while effective temperature is controlled by adjusting the entropy at the bottom of 3D models. It is therefore difficult to match an accurate effective temperature. 

\subsection{Computation of patched models} 

For each 3D model, both a patched (hereafter PM) and an unpatched model (hereafter UPM) have been constructed. A patched model is a model computed using a 1D stellar evolutionary code in which we replaced the surface layers with the stratification obtained by horizontally averaging a 3D model computed with a R-MHD code. The fully 1D model is called an unpatched model. The construction of PM and UPM has been widely described in \citet{Trampedach1997,Samadi2007,Samadi2008,Sonoi2015,Jorgensen2017}. The 1D counterparts of 3D hydrodynamical models have been obtained using the 1D stellar evolutionary code CESTAM \citep{Morel1997,Marques2013} by tuning the age (or the central temperature for advanced stages), the total stellar mass $M$, and the mixing length parameter $\alpha_{\rm MLT}$ in order to match the effective temperature, the surface gravity and the temperature at the bottom of the 3D model, located just below the superadiabatic region. We chose to remove the first four bottom layers and the last top layer of the 3D hydrodynamical model to be sure to remove any numerically induced errors and that the patching point is deeply inside the adiabatic region, which has been shown to be a condition for a obtaining reliable PM \citep{Jorgensen2017}.

The 1D models use the equation of state, and opacities given by OPAL2005 \citep{Rogers2002,Iglesias1996} and implement standard mixing-length theory \citep{Bohm-Vitense1958} with no overshoot. We ignore diffusion processes, rotation and turbulent pressure. The atmosphere is computed using the Eddington approximation. The helium abundance in 1D models is set to the one used in 3D models.

Finally, we note that for some 3D models one can find a degenerate solution for the corresponding 1D model: we could patch either a PMS or a sub-giant model. We opted for evolved models since they corresponds to stars in which solar-like oscillations are observed so far. However, when the evolved models are too old (older than the age of the Universe) we kept the PMS model, except if lying on the Hayashi track.

Table \ref{table:3d_param} also summarizes the stellar parameters of both UPM and PM together with relative radius differences $R_{\rm PM} / R_{\rm UPM} - 1$. Our set of models covers a wide portion of the Hertzsprung-Russel Diagram for intermediate mass stars. We note that our patched models with metallicity $[\rm Fe/H]=+0.5$ only have $\log g = 4.0$. Indeed, 3D models from the CIFIST grid with $[\rm Fe/H] = +0.5$ were only available for $\log g \geq 4.0$. In addition, 3D models with $\log g \gtrsim 4.5$ are located below the main sequence diagonal, and therefore it is impossible to find a 1D model matching their characteristics (with the physical ingredients we used). Thus, a large portion of our initial $[\rm Fe/H] = +0.5$ 3D hydrodynamical models were not suitable for our purposes.

\subsection{Computation of oscillation frequencies}

In this work, we consider only structural effects and an adiabatic treatment of the oscillations. The frequencies are computed using the ADIPLS code \citep{Christensen-Dalsgaard2011}  for both UPM and PM by assuming the gas $\Gamma_1$ approximation, which assumes that the relative Lagrangian perturbations of gas pressure and turbulent pressure are equal \citep{Rosenthal1999,Sonoi2017}. Besides this distinction in the treatment of $\Gamma_1$ entering the calculation of the model frequencies, we emphasize that the frequency differences studied in this work are only emerging from structural effects. Therefore, it should be emphasized that the frequency differences studied in this work concern only purely structural effects. We have checked that we recovered the previous results of \cite{Sonoi2015} for the solar metallicity. For the sake of simplicity, we mainly focussed on the surface effect affecting radial modes: non-radial modes exhibit a mixed behaviour that would make our analysis more complex (however, see Sect. \ref{subsubsection:mixed_modes} for a discussion).

\section{Influence of metallicity}

Until now, surface effects have always been studied assuming a solar metallicity. Corrections depend only on $T_{\rm eff}$ and $\log g$ such as the power law proposed by \citet{Kjeldsen2008}, cubic and combined inverse--cubic laws \citep{Ball2014}, or a modified Lorentzian \citep{Sonoi2015}. This section is intended to motivate the investigation of the dependence of the surface effect on metallicity. 

\begin{figure}
    \centering
    \includegraphics[width=\hsize]{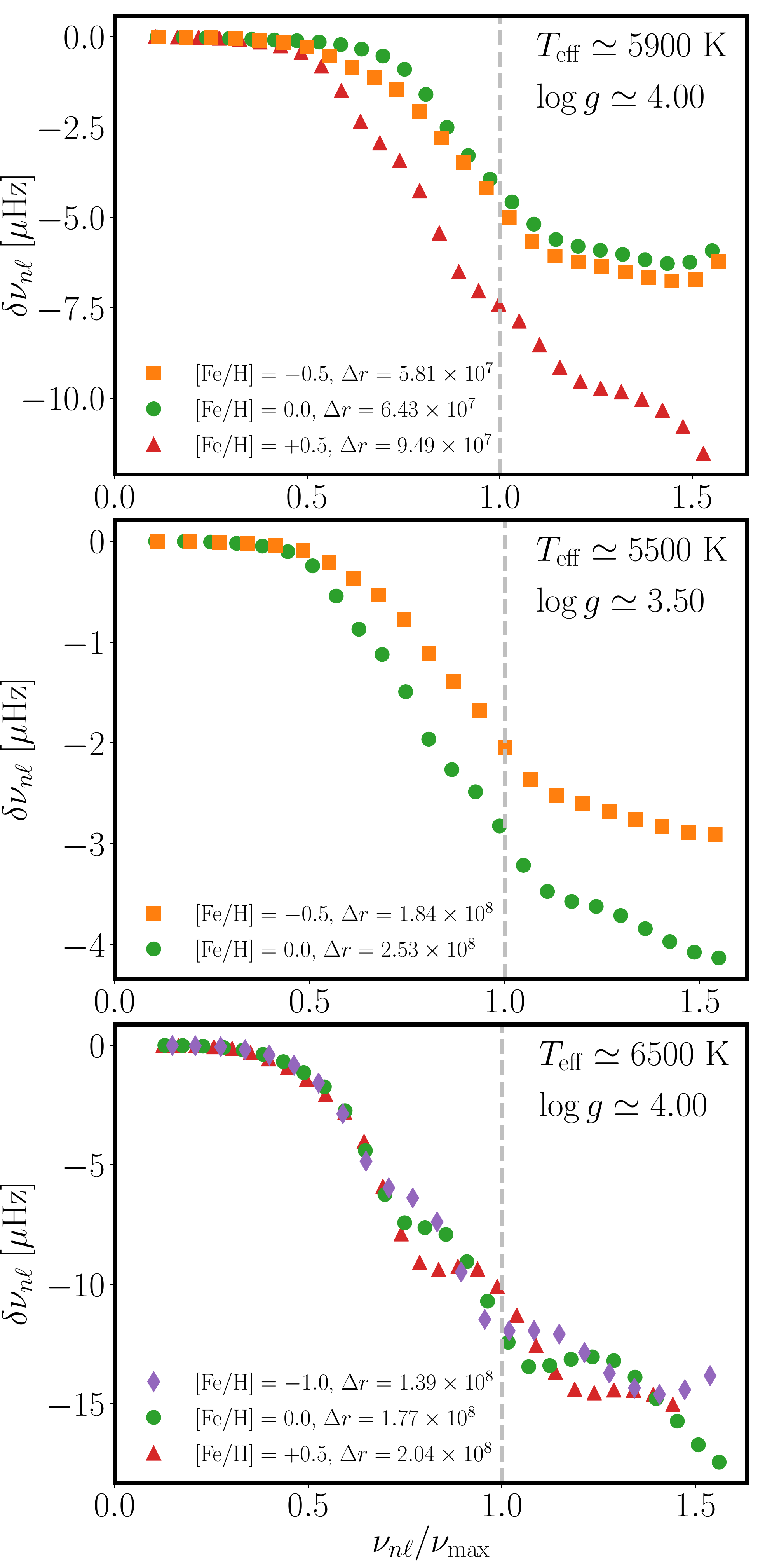}
    \caption{Frequency differences of PM vs. UPM for radial modes with frequencies less than the acoustic cut-off frequency $\nu_{\rm ac} = c / (4 \pi H_p)$. Frequencies on the abscissa are normalized by the $\nu_{\rm max} $ of each models. \textsc{top panel:} Models J*: $T_{\rm eff} \simeq 5900 \mbox{K}$ and $\log g = 4.0$. \textsc{middle panel:} Models M*: $T_{\rm eff} \simeq 5500 \mbox{K}$ and $\log g = 3.5$. \textsc{bottom panel}: Models C*: $T_{\rm eff} \simeq 6500 \mbox{K}$ and $\log g = 4.0$.}
    \label{fig:influ_met_dnu}
\end{figure}

\subsection{Qualitative influence of metallicity on frequency differences}
\label{subsection:s3_qualitative}

We begin this section by quickly describing the effects of a change of metallicity on the frequency residuals. Frequency differences are induced by the surface elevation between PM and UPM due to turbulent pressure, which extends the size of the resonant cavities and therefore decreases the mode frequencies for PM, leading to negative frequency differences $\delta \nu = \nu_{\rm PM} - \nu_{\rm UPM}$. 

Up to now, only the influence of effective temperature and surface gravity on surface effects have been studied. However, the abundance of heavy elements has a strong impact on opacity and hence on the convective flux imposed by a change in the radiative flux. In turn, a change in the convective flux leads to a change of convective velocity and therefore a change of turbulent pressure and finally it changes the location of the surface. We mention here that metallicity also has an effect on gas pressure, through the mean molecular weight $\mu$, which varies in the opposite direction of the turbulent pressure and therefore counteracts its effect. Finally, while mechanisms by which a change of metallicity can act on the surface effect are known, those mechanisms are too intricate to identify the resulting effect on the variations of surface term without a deeper analysis as will be demonstrated in the following (see Section \ref{section:s4}).

Figure \ref{fig:influ_met_dnu} shows the (purely structural) frequency differences for three groups of models that have approximately the same effective temperature and surface gravity. The discrepancies in $\nu_{n\ell}$ between two models appear at relatively low frequencies and generally increase towards high frequencies. As for finding a general trend of the evolution of the surface effect against the metallicity, it seems from Fig. \ref{fig:influ_met_dnu} no such trend exists: in the top panel, frequency differences, at $\nu_{\rm max}$ for instance, slightly decrease from $[\rm Fe/H] = -0.5$ to $0.0$ and then are much higher for the $[\rm Fe/H] = +0.5$ model. In the middle panel, the frequency residual at $\nu_{\rm max}$ significantly increases from $[\rm Fe/H] = -0.5$ to $0.0$. Finally, in the bottom panel, very little variations at $\nu_{\rm max}$ can be noticed from one composition to an other. However, the variation of the frequency differences seems to follow closely the variations of the elevation of the stellar surface between UPM and PM: 
\begin{equation}
    \Delta r \equiv R_{\rm PM} - R_{\rm UPM}.
    \label{eq:Deltar}
\end{equation}
The slight disagreement between $[\rm Fe/H] = 0.0$ and $-0.5$ in the top panel may be explained by the large dispersion in effective temperature.

\subsection{Effect of the elevation on the frequency differences}
\label{subsection:s3_elevation}

To gain some insight into the influence of metallicity on surface effect, we tried to scale the normalized frequency differences at $\nu_{\rm \rm max}$ for our set of models. This is a necessary step to allow an estimate of the surface effect correction parameters (see Sect.~\ref{section:s4}). Thus, let us start with the perturbative approach as adopted by \citet{Christensen-Dalsgaard1997}\citep[see also][]{Goldreich1991,Balmforth1996}. The authors show that the frequency difference can be well approximated by 
\begin{align}
    \frac{\delta \nu}{\nu} &= \int_0^R\left[\tilde{K}_{c^2, v}^{n\ell} \frac{\delta_m c^2}{c^2} + \tilde{K}_{v, c^2}^{n\ell} \frac{\delta_m v}{v}\right] {\rm d} r\\
    &\simeq \int_0^R \tilde{K}_{v,c^2}^{n\ell} \frac{\delta_m v}{v} {\rm d} r \, ,
    \label{eq:deltanu_kernel}
\end{align}
where $c$ is the adiabatic sound speed, the variable $v$ is defined by $v=\Gamma_1/c$, $\tilde{K}_{c^2, v}^{n\ell}$ and $\tilde{K}_{v, c^2}^{n\ell}$ are the kernels that can be determined from eigenfunctions, $\delta_m c^2$ and $\delta_m v$ are the Lagrangian differences of $c^2$ and $v$, respectively, at fixed mass. 

\cite{Rosenthal1999} further approximated the frequency differences for radial modes, based on the expression of $\tilde{K}_{v,c^2}^{n\ell}$ and using a first-order asymptotic expansion for the eigen-function, by
\begin{equation}
    \frac{\delta \nu}{\nu} \simeq \frac{\Delta \nu \Delta r}{c_{\rm ph}} \, ,
    \label{eq:deltanu_elevation}
\end{equation}
where $\Delta \nu$ is the asymptotic large frequency separation, $\Delta r$ is the previously defined elevation, and $c_{\rm ph}$ the photospheric sound speed (see Appendix \ref{appendix:app1} for a demonstration of this relation).

\begin{figure}
    \includegraphics[width=\hsize]{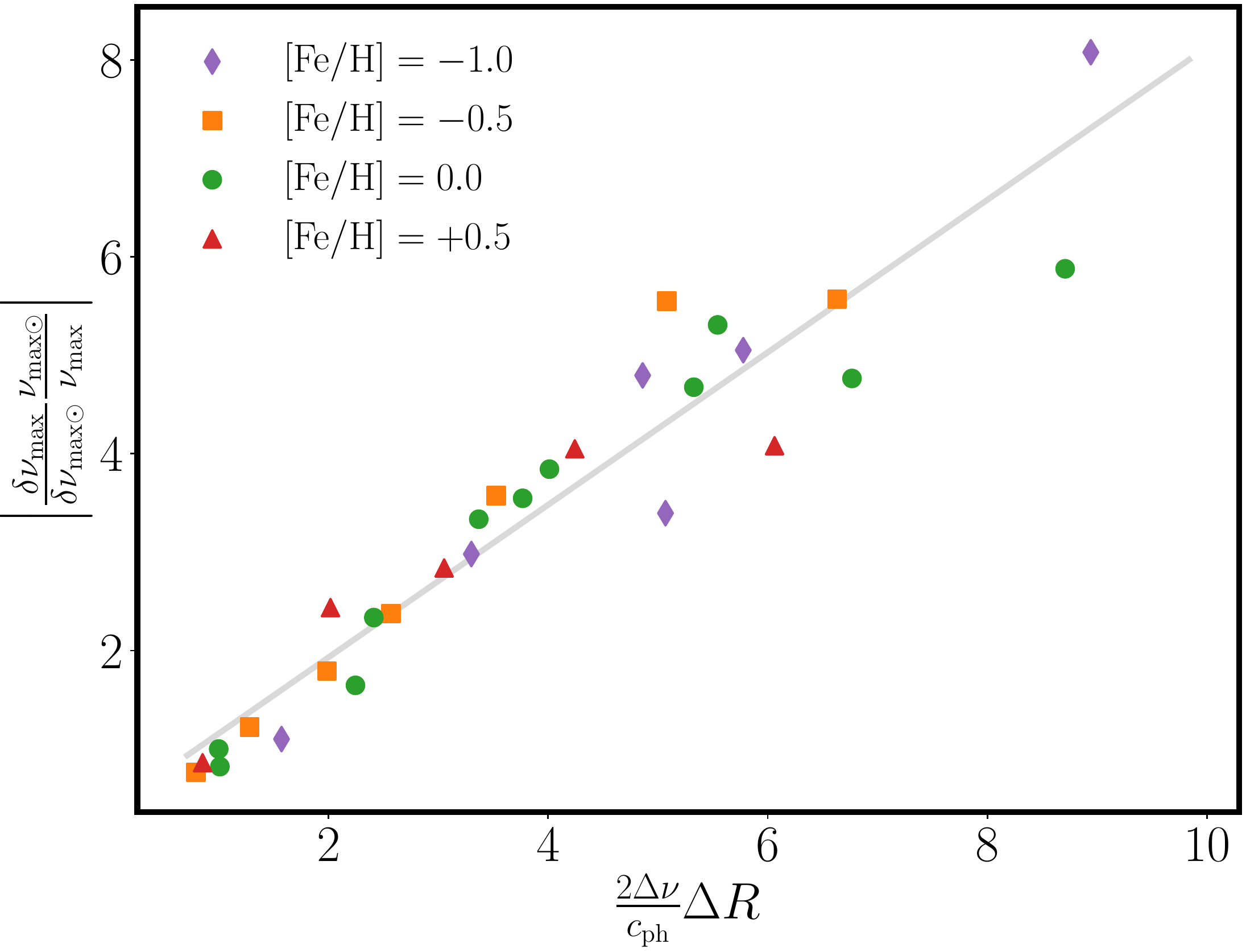}
    \includegraphics[width=\hsize]{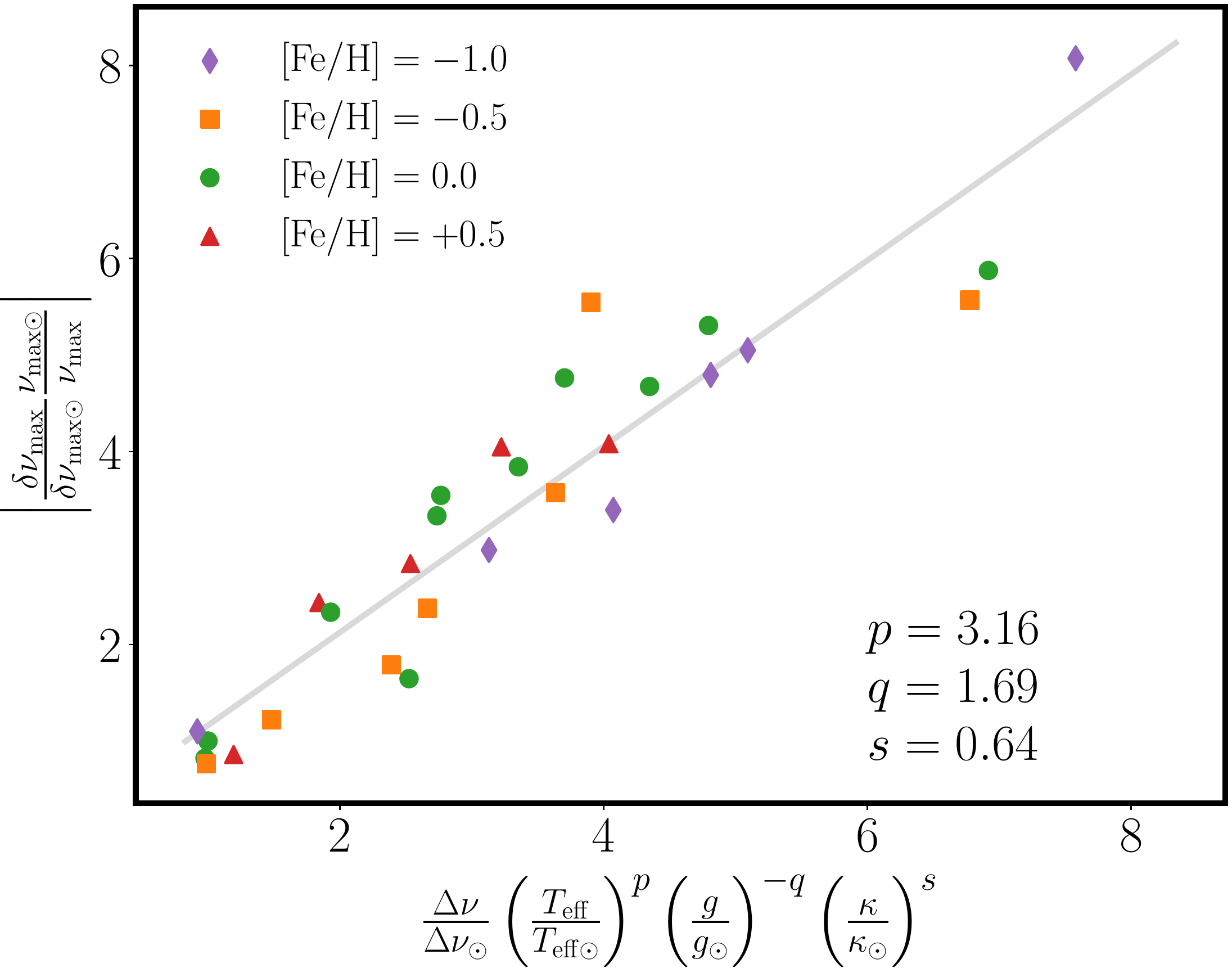}
    \caption{Frequency differences scaled by the frequency, taken at $\nu_{\rm max}$ against a scaling relation given by \citet{Rosenthal1999} (top panel) (with $\sigma = 0.89$) and a scaling relation given by Eq. \eqref{eq:z1} where powers are left free (bottom panel) (with $\sigma_\kappa = 0.63$).}
    \label{fig:dnu_nu_rosenthal}
\end{figure}

This relation has been previously tested by \cite{Sonoi2015} at solar metallicity using surface effect derived from a grid of 3D numerical simulations. It turns out that Eq.~\eqref{eq:deltanu_elevation} reproduces the overall scale of the surface effect (such as in Fig. \ref{fig:dnu_nu_rosenthal} were the surface effect is considered at $\nu_{\rm max}$) for a set of models. It is thus necessary to determine whether this relation holds for models with a non-solar metallicity. To this end, we have compared frequency residuals at $\nu=\nu_{\rm max}$ given by Eq.~\eqref{eq:deltanu_elevation} as shown in Fig. \ref{fig:dnu_nu_rosenthal} (top panel). There is still a good agreement between the frequency differences and the approximated expression given by Eq.~\eqref{eq:deltanu_elevation}. Moreover, it appears that the frequency differences are dominated by the surface elevation $\Delta r$. To understand the link to metallicity, it is thus necessary to go a step further and to investigate the relation between surface elevation and metallicity. 

\subsection{Scaling law for the frequency differences}
\label{subsection:s3_freq_scaling}

In this section, we aim to determine a relation between frequency differences at $\nu_{\rm max}$ and global parameters of the models. First, as shown in the previous section, there is no clear trend between the surface term and metallicity. Indeed, at constant metallicity and considering our rather large range of effective temperatures and surface gravities, the dominant opacity mechanisms are not the same from a model to an other for instance, the opacity at the surface is dominated by the negative hydrogen ions for $T_{\rm eff} \lesssim 5000~\mbox{K}$. Therefore, the relation between $\delta \nu / \nu$ and $Z$ is non-trivial. To overcome this problem, we directly consider the Rosseland mean opacity at the photosphere instead of the metallicity as a global parameter in addition to the effective temperature and to the surface gravity (in the following, the photosphere is defined as the radius at which $T = T_{\rm eff}$).

Let us begin by considering the elevation in Eq. \eqref{eq:Deltar} which must be expressed as a function of these global parameters. Using the hydrostatic equilibrium equation, it reads
\begin{equation}
    \Delta r = \int_0^{R_{\rm PM}} H_p^{\rm PM} \frac{{\rm d}p_{\rm tot}}{p_{\rm tot}} 
    - \int_0^{R_{\rm UPM}} H_p^{\rm UPM} \frac{{\rm d}p_{\rm g}}{p_{\rm g}} \, , 
    \label{eq:hydrostatic_elevation}
\end{equation}
where $H_p^{\rm PM}$ and $H_p^{\rm UPM}$ are the pressure scale heights at the photosphere associated with the patched and unpatched models, $p_{\rm tot}$ is the total pressure such as $p_{\rm tot}=p_{\rm turb}+p_{\rm g}$ with $p_{\rm turb}$ and $p_{\rm g}$ the turbulent and gas pressure, respectively. Further assuming that $H_p^{\rm PM} \simeq H_p^{\rm UPM}$ and $p_{\rm turb} / p_{\rm tot} \ll 1$ one can approximate Eq. \eqref{eq:hydrostatic_elevation} by
\begin{equation}
    \Delta r \simeq H_p^{\rm PM} \frac{p_{\rm turb}}{p_{\rm tot}} \,.
\end{equation}

Finally, since the pressure scale-height scales as $T_{\rm eff}/g$, the elevation scales as $\Delta r \propto {(T_{\rm eff}p_{\rm turb})}/{(gp_{\rm g})}$.

To go further, we need to find an expression for $p_{\rm turb}/ p_{\rm g}$. Near the photosphere, the turbulent pressure can be written as 
\begin{equation}
    p_{\rm turb} = \rho v_{\rm conv}^2,
    \label{eq:pt}
\end{equation}
where $v_{\rm conv}$ is the vertical component of the convective velocity. We now need an expression for this velocity and for the density. Assuming a standard Eddington grey atmosphere, the optical depth is approximated by $\tau = H_p \rho \kappa$, and in the Eddington approximation, we have $\tau = \nicefrac{2}{3}$ at the bottom of the photosphere. Then, and accordingly:
\begin{equation}
    \rho \propto \frac{g}{T_{\rm eff} \kappa}.
    \label{eq:rho}
\end{equation}
As for finding an expression for $v_{\rm conv}$, we note that $F_{\rm tot} = F_{\rm rad} + F_{\rm conv}$, with $F_{\rm rad}$ and $F_{\rm conv}$ the radiative and convective component of the total energy flux respectively. The convective flux is proportional to the kinetic energy flux (as shown for instance within the MLT framework). Then,
\begin{equation}
    \rho v_{\rm conv}^3 \propto T^4_{\rm eff}\left(1 + \displaystyle{\frac{F_{\rm rad}}{F_{\rm conv}}}\right)^{-1}.
    \label{eq:Ftot}
\end{equation}
The ratio $F_{\rm rad} / F_{\rm conv}$ is assumed to remain nearly constant from one model to an other. Therefore, $v_{\rm conv}$ finally reads, 
\begin{equation}
    v_{\rm conv}^2 \propto \frac{T_{\rm eff}^{\nicefrac{8}{3}}}{\rho^{\nicefrac{2}{3}}}.
    \label{eq:vconv}
\end{equation}

Inserting the expressions of Eq. \eqref{eq:rho} and \eqref{eq:vconv} into Eq. \eqref{eq:pt} leads to:
\begin{equation}
    p_{\rm turb} \propto \left(\frac{T_{\rm eff}}{T_{\rm eff \odot}}\right)^{\nicefrac{7}{3}}\left(\frac{g}{g_\odot}\right)^{\nicefrac{1}{3}}\left(\frac{\kappa}{\kappa_\odot}\right)^{-\nicefrac{1}{3}},
\end{equation}
where $\kappa_\odot = 0.415~\mbox{cm}^2\cdot\mbox{g}^{-1}$. From the perfect gas law for, $p_{\rm g} \propto \rho T_{\rm eff}$ and using Eq. \eqref{eq:rho}, we can rewrite $\Delta r$ as
\begin{equation}
    \Delta r \propto \left(\frac{T_{\rm eff}}{T_{\rm eff \odot}}\right)^{\nicefrac{10}{3}}\left(\frac{g}{g_\odot}\right)^{-\nicefrac{5}{3}}\left(\frac{\kappa}{\kappa_\odot}\right)^{\nicefrac{2}{3}}.
\end{equation}

Replacing $\Delta r$ into Eq. \eqref{eq:deltanu_elevation} one finally obtains the following estimate:

\begin{equation}
\frac{\delta \nu}{\nu} \propto \frac{\Delta \nu}{\Delta \nu_\odot} \left(\frac{T_{\rm eff}}{T_{\rm eff \odot}}\right)^{\nicefrac{17}{6}}\left(\frac{g}{g_\odot}\right)^{-\nicefrac{5}{3}}\left(\frac{\kappa}{\kappa_\odot}\right)^{\nicefrac{2}{3}} \equiv z_1.
\label{eq:z1}
\end{equation}

This expression provides us with a simple relation between the frequency differences and the global parameters. The dependence on the metallicity is embedded into the Rosseland mean opacity. We note that it is possible to go further and to explicitly introduce the metallicity. For instance, in the vicinity of the solar effective temperature and gravity, the opacity is dominated by the $H^-$ so that $\kappa \propto \rho^{-\nicefrac{1}{2}} \, T_{\rm eff}^9 Z$. However, given the wide range of effective temperatures and surface gravities of our grid of models, it is more relevant to keep the Rosseland mean opacity at $T=T_{\rm eff}$ (surface opacity) as a global parameter. Indeed, the Rosseland mean opacity is a quantity available in any 1D stellar evolutionary code.     

Then, using Eq.~(\ref{eq:z1}) as a guideline, we performed a fit where the powers of the temperature ($p$), gravity ($q$), and opacity ($s$) have been  adjusted at $\nu=\nu_{\rm max}$ for each model. Figure \ref{fig:dnu_nu_rosenthal}, bottom panel, displays the result. This figure shows a very good agreement between exponents derived in Eq. \eqref{eq:z1} and the one actually obtained using our simulations. Consequently this scaling can be used to provide a physically-grounded values for the parameters of the empirical correction function of the surface effect. Finally, we note that using the opacity instead of the metallicity allows us to take a detailed mixture into account. 

In addition to our crude approximations, a possible source of discrepancies between values predicted by Eq. \eqref{eq:z1} and the one calculated can be that we did not fix the helium abundance from one model to the other when varying the metallicity. The changing helium abundances have an impact both on the evolution of the model and on its opacity at the surface. However, the helium abundances $[\rm He/H]$ range between $- 5.8\times 10^{-3}$ and $+1.2\times10^{-2}$ and should be a negligible source of uncertainty. A final source of error comes from the method we used to average the 3D stratifications. Indeed, since the Rosseland opacity is involved Eq. \eqref{eq:z1}, it would be more precise to patch the models using a stratification averaged against the Rosseland optical depth instead of the actual geometrically averaged stratification, but this is beyond the scope of this paper and will be investigated in a forthcoming work.

\section{Surface-effect corrections}

\label{section:s4}

\begin{figure}
    \centering
    \includegraphics[width=\hsize]{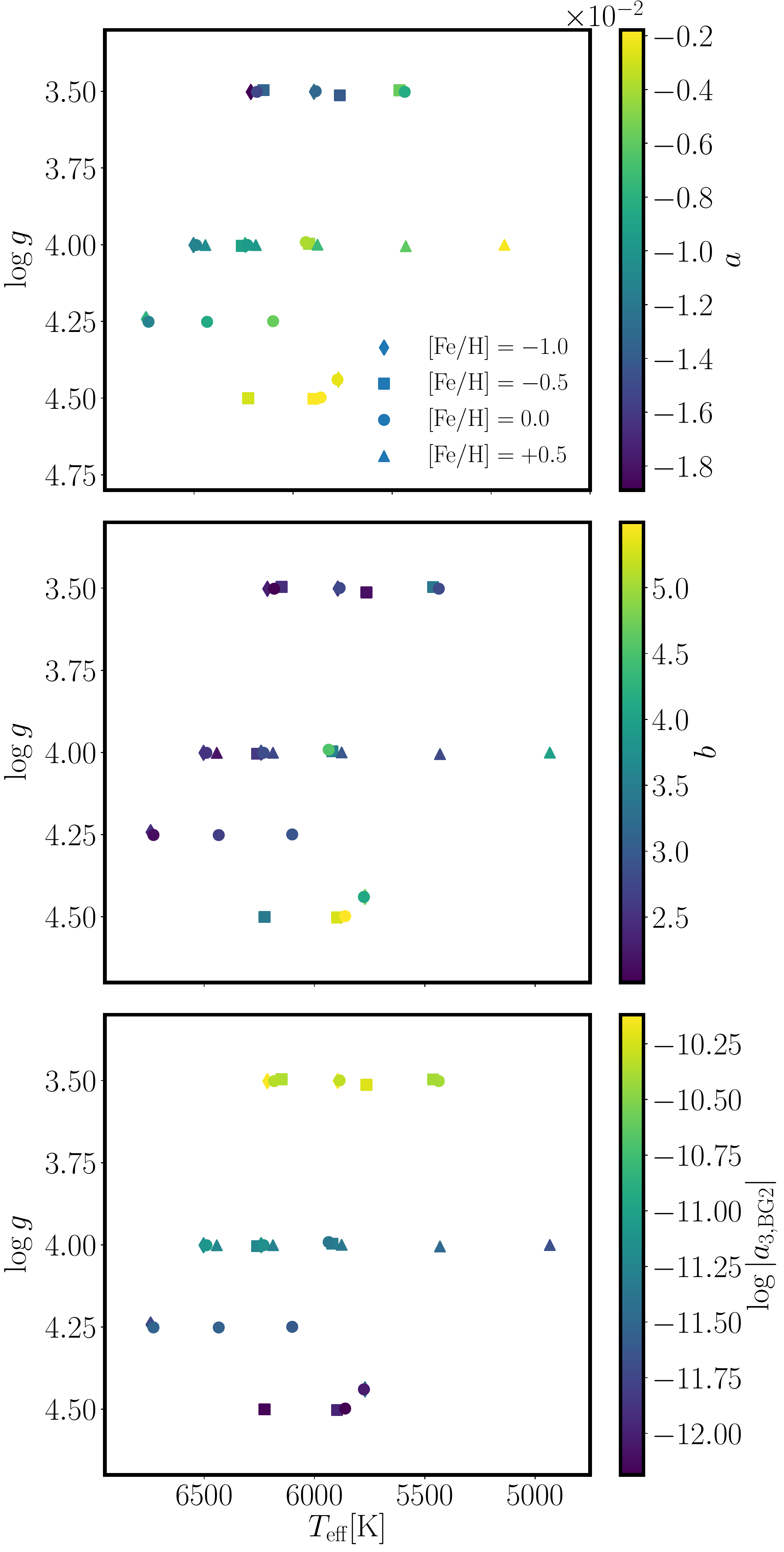}
    \caption{Parameters $a$ (top panel), $b$ (middle panel) and $a_{3\rm , BG2}$ (bottom panel) across the Kiel diagram from K08 and BG2. The symbols refer to $\rm [Fe/H] = -1.0$ (diamond), $\rm [Fe/H] = -0.5$ (square), $\rm [Fe/H] = +0.0$ (circle) and $\rm [Fe/H] = +0.5$ (triangle).}
    \label{fig:a_b_b4}
\end{figure}

A handful of empirical functions have been suggested to perform a posteriori corrections on the modelled frequencies. After having given a theoretical background that explains variations of $\delta \nu / \nu$, we considered the most commonly used correction models to study the evolution of the related free parameters as a function of effective temperature, surface gravity, and surface opacity. This is intended to provide constraints on those parameters and thus to provide physically-grounded values for use on seismic observations.

\subsection{Empirical functions for correcting modelled frequencies}

\subsubsection{\citet{Kjeldsen2008} power law}

\citet{Kjeldsen2008} proposed a power law which  was found to match the frequency differences obtained between the observed and modelled solar frequencies:
    \begin{equation}
        \frac{\delta \nu}{\nu_{\rm max}} = a \left[\frac{\nu_{\rm PM}(n)}{\nu_{\rm max}}\right]^b \, , 
        \label{eq:kjeldsen}
    \end{equation}
where $a$ and $b$ are the parameter to be adjusted. They found $a = -4.73$ and $b = 4.9$ for their model of the Sun by matching a subset of nine radial modes centred on $\nu_{\rm max}$. 

\citet{Kjeldsen2008} provided a method to correct the frequency for a star similar to the Sun without having to calibrate $b$. Let us assume we want to model a star with near solar global parameters and we want to constrain our model using the individual frequencies. The radial mode frequencies spectrum of our best model which include a surface term are denoted $\nu_{i, \rm best}$ and the frequencies of solar radial modes for the same order are denoted $\nu_{i, \rm ref}$. Then, \citet{Kjeldsen2008} proposed that the frequencies can be linked, to a good approximation, by $\nu_{i, \rm best} \simeq r\nu_{i, \rm ref}$, using the proportionality factor $r$ between mean densities of both models: $\bar{\rho}_{\rm best} = r^2\bar{\rho}_{\rm ref}$. Using this relation and the large separations of both models, they provided a way to obtain $a$ and $b$. Further assuming $b$ constant (the value of which depends of the physical ingredients used in the model), they derived a value for $a$ for a set of theoretical models close to the Sun.

This power law has been widely used since and many authors \citep[eg.][]{Metcalfe2009,Bedding2010} have used a constant value for $b$ (not necessarily 4.90 though) derived from solar frequency measurements. Keeping $b$ constant is often necessary in the case for which observations do not provide enough constraints to adjust it. However, using the solar value leads to a bad correction if the modelled star is too different from the Sun \citep[eg.][]{Kallinger2010}. Furthermore, $b_\odot$ depends on the input physics. Otherwise, $b$ can be considered as a variable parameter in the modelling and therefore significantly improve the correction. Different models of the star HD 52265 have been compared by \citet{Lebreton2014} using various input physics and found approximatively the same predicted age models when either frequency ratios \citep{Roxburgh2003} or individual corrected frequencies were used as constraints. The age dispersion was slightly higher with models constrained by individual corrected frequencies ($\sim \pm 9.5 \%$) and using uncorrected individual frequencies lead to ages 40\% larger \citep{Lebreton2014II}.

In the following, we have studied two versions of this parametric function. The first, adjusted on the whole radial mode frequency spectrum for frequency less than the acoustic cut-off frequency, will be referred to as K08. The second, adjusted on a reduced frequency interval $0 < \nu/\nu_{\rm max} < 1.05$ is refered to as K08r (see Fig. \ref{fig:dev_plot} and Appendices \ref{appendix:2} and \ref{appendix:3}).

\begin{figure*}
    \centering
    \includegraphics[width=0.98\hsize]{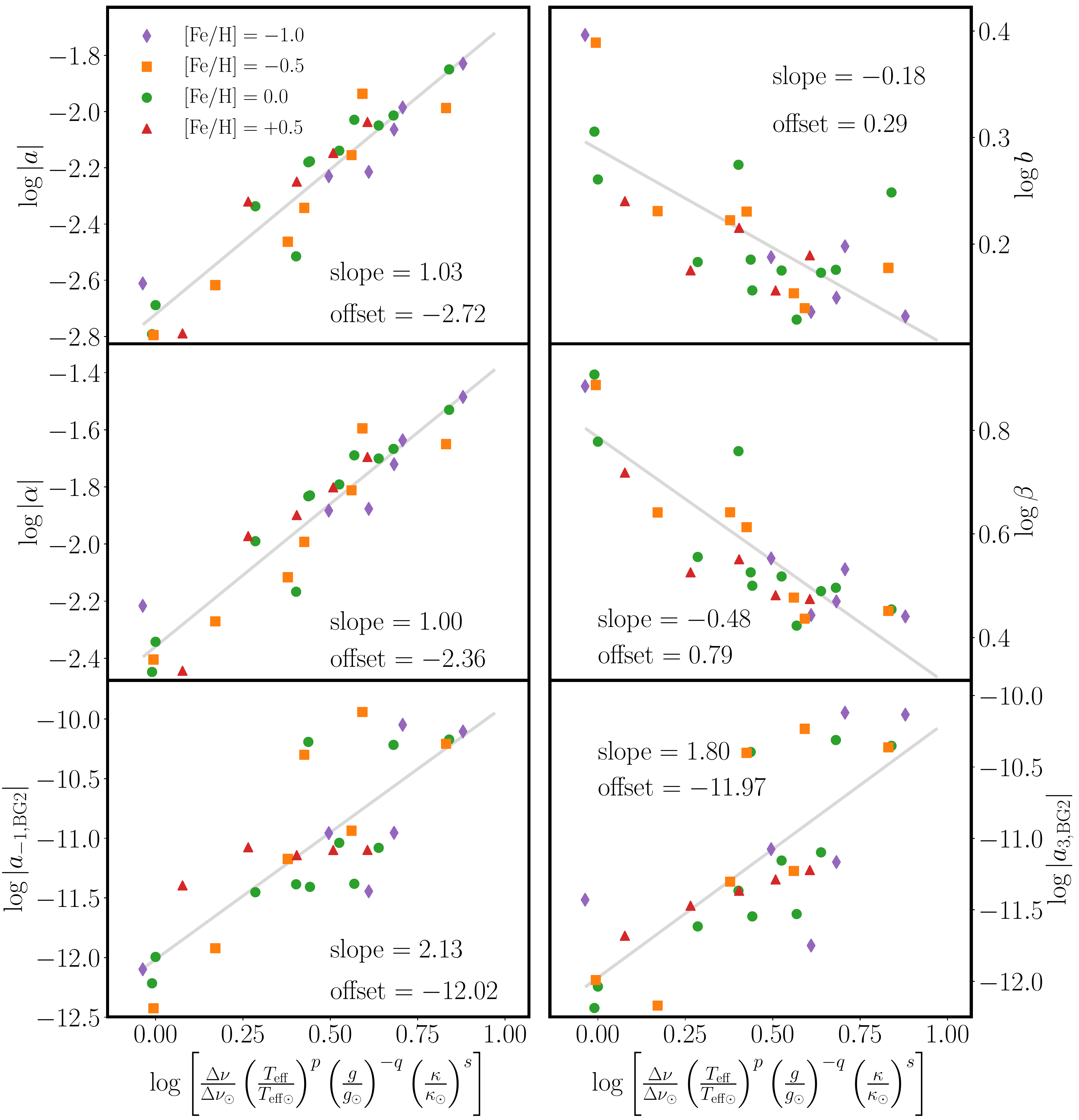}
    \caption{Logarithms of K08 (first row), S15 (second row) and BG2 (third row; $\log |a_{3\rm , BG1}|$ behaves almost identically to $\log |a_{3\rm , BG2}|$) fitting coefficients as a function of $\log z_1$ (Eq. \eqref{eq:z1}): $\log f = {\rm slope} \times \log z_1 + {\rm offset}$. Purple dots (resp. yellow, green and red) corresponds to $[\rm Fe/H] = -1.0$ (resp. $-0.5$, $0.0$ and $+0.5$).}
    \label{fig:coeff_fit}
\end{figure*}

\subsubsection{\citet{Ball2014} cubic and combined inverse-cubic laws}
\label{subsubsection:ball}

\citet{Ball2014} suggested a new function to correct frequency differences. It is partially based on the early work by \cite{Gough1990} (for the cubic part). They accounted for two leading effects introducing systematic errors in the theoretical computation of the frequency spectrum: the modification of the sound speed caused by a magnetic field concentrated into a filament by convective motions, causing a frequency shift scaling as $\nu^3 / E$ \citep{Libbrecht1990}, $E$ being the normalized mode inertia; and the modification of the pressure scale height caused by a poor description of convection, inducing a frequency shift scaling as $\nu^{-1} / E$. This correction funtional has the  advantages of being independent of a solar calibration and including a dependence on the normalized mode inertia which allows us to correct non-radial modes, without the need of re-scaling their frequency differences. Because of this, they suggested a cubic correction taking only into account the dominant effect and a combined inverse-cubic correction including the perturbation.

The cubic correction (in the following BG1) is defined by 
\begin{equation}
    \frac{\delta \nu}{\nu_{\rm max}} = \frac{a_{3\rm , BG1}}{E} \left(\frac{\nu}{\nu_{\rm max}}\right)^3 \, , 
    \label{eq:cubic}
\end{equation}
and the combined inverse-cubic correction (in the following BG2) is 
\begin{equation}
    \frac{\delta \nu}{\nu_{\rm max}} = \frac{1}{E}\left[a_{-1\rm , BG2} \left(\frac{\nu}{\nu_{\rm max}}\right)^{-1} + a_{3\rm , BG2} \left(\frac{\nu}{\nu_{\rm max}}\right)^3\right] \, , 
    \label{eq:combined}
\end{equation}
where $E$ is the normalized mode mass:
\begin{equation}
    E = \frac{4 \pi \int_0^R\left[|\xi_{\rm r}(r)|^2 + \ell(\ell + 1)|\xi_{\rm h}(r)|^2 \right]\rho r^2 {\rm d}r}{M\left[|\xi_{\rm r}(R)|^2 + \ell(\ell + 1)|\xi_{\rm h}(R)|^2 \right]},
    \label{eq:mode_mass}
\end{equation}
where $R$, $M$, and $\rho$ are respectively the photospheric radius, mass and density of the star, and $\xi_{\rm r}$ and $\xi_{\rm h}$ are the radial and the horizontal component of the displacement of an eigenmode of degree $\ell$. $a_{3\rm , BG1}$, $a_{-1\rm , BG2}$, and $a_{3\rm , BG2}$ are the parameters to be adjusted. They used the acoustic cut-off frequency $\nu_{\rm c}$ instead of $\nu_{\rm max}$ in order to normalize their fitting parameters: it only results in a modification of $a_{-1}$ and $a_3$ and does not change the law itself.

\subsubsection{\citet{Sonoi2015} modified Lorentzian}

\begin{table*}
    \caption{Prescriptions for the fitting coefficients involve in the empirical relations K08, K08r, S15, BG1, BG2. The prescriptions are written: $\log c_0 = c_1 \log \Delta\nu/\Delta\nu_\odot + c_2 \log T_{\rm eff}/T_{\rm eff\odot} + c_3 \log g/g_\odot + c_4 \log \kappa/\kappa_\odot +c_5$. Our solar values are computed from model Am00. $\Delta \nu_\odot = 137 ~\rm [\mu Hz]$, $T_{\rm eff\odot} = 5776~\rm [K]$, $g_\odot = 27511 ~\rm[cm\cdot s^{-2}]$, and $\kappa_\odot = 0.415~\rm[g\cdot cm^{-2}]$.}
    \label{table:prescriptions}     
    \centering 
    {\renewcommand{\arraystretch}{1.2}
    \begin{tabular}{l l | c c c c c}       
        \hline\hline                   
Law                   & $\log c_0      $ & $c_1$      & $c_2$     & $c_3$     & $c_4$     & $c_5$   \\
\hline
\multirow{2}{*}{K08}  & $\log |a|      $ & $1.03  $ & $3.26$    & $-1.75$   & $0.655$   & $-2.72 $  \\ 
                      & $\log b        $ & $-0.185$ & $-0.584$  & $0.313$   & $-0.117$  & $0.289$   \\
\hline
\multirow{2}{*}{K08r} & $\log |a|      $ & $1.08  $ & $3.39$    & $-1.82 $  & $0.683$   & $-2.65 $  \\
                      & $\log b        $ & $-0.387$ & $-1.22$   & $0.655$   & $-0.246$  & $0.647$   \\
\hline
\multirow{2}{*}{S15}  & $\log |\alpha| $ & $0.999 $ & $3.15$    & $-1.69 $  & $0.635$   & $-2.36 $  \\
                      & $\log \beta    $ & $-0.477$ & $-1.51$   & $0.808$   & $-0.303$  & $0.787$   \\
\hline
BG1                   & $\log |a_3|    $ & $1.93  $ & $6.09$    & $-3.26 $  & $1.22$    & $-11.9$   \\
\hline
\multirow{2}{*}{BG2}  & $\log |a_{-1}|$  & $2.13  $ & $6.72$    & $-3.6  $  & $1.35$    & $-12$     \\
                      & $\log |a_3|    $ & $1.8   $ & $5.67$    & $-3.04 $  & $1.14$    & $-12$     \\
\hline
    \end{tabular}}
\end{table*}

The final function to be introduced was a modified Lorentzian \citep{Sonoi2015} that was found to better correct the surface effect derived from the 3D simulations at high frequency. It reads
\begin{equation}
    \frac{\delta \nu}{\nu_{\rm max}} = \alpha \left[1 - \frac{1}{1+\left(\frac{\nu_{\rm PM}}{\nu_{\rm max}}\right)^\beta}\right] \, ,
    \label{eq:lorentz}
\end{equation}
where $\alpha$ and $\beta$ parameters are to be determined. When $\nu_{\rm PM}/\nu_{\rm max} \ll 1$ we get back to \citet{Kjeldsen2008} law. When $\nu_{\rm PM} = \nu_{\rm max}$, ${\delta \nu}/{\nu_{\rm max}} = \alpha / 2$. Therefore, $a$ and $\alpha$ are directly linked to ${\delta \nu}/{\nu}$ given by Eq. \eqref{eq:z1}, which gives physical justification for its variations. In the following, we will refer to this correction law as S15.

A recent comparison of the above correction laws has been performed by \citet{Ball2017} on six sub- and red giants from the \textit{Kepler} Input Catalog. We note that since these are evolved stars, they display mixed-modes which have their frequency residuals off the general trend of radial p-modes frequency residuals. This should have consequences on the quality of the correction. They computed stellar models matching their six stars constrained by the effective temperature, the metallicity and the individual frequencies. They tested five correction relations: BG1, BG2, S15, and K08 with power $b = 5.0$ calibrated with a solar model computed with their own input physics and K08 with power $b$ left free.

\citet{Ball2017} found no correction to be clearly superior than the others for all stars. However BG2 and then BG1 performed slightly better than the others, followed by the free power law, S15 and finally K08. S15 was shown to poorly correct high frequencies and the K08 with $b = 5.0$ correction gave worse results than no correction for their of their stars. We present very similar conclusions in the following.

\begin{figure*}
    \centering
    \includegraphics[width=\hsize]{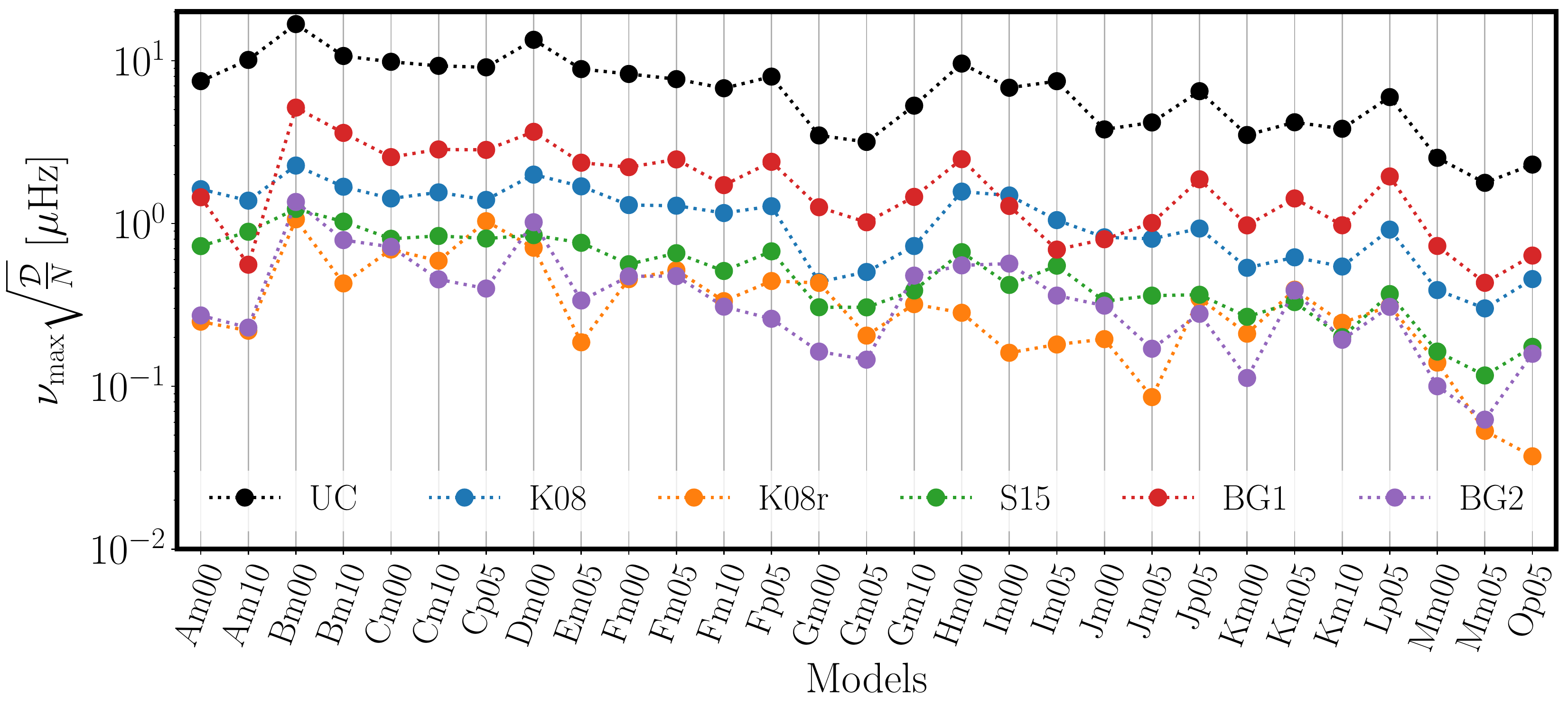}\\
    \caption{Root-mean frequency differences $\nu_{\rm max}\sqrt{\mathcal{D}/N}$ of radial modes after correction for each empirical law and for each model. It shall be noticed that the deviation for K08r is computed only on the range $0 < \nu/\nu_{\rm max} < 1.05$. Black dots corresponds to uncorrected frequencies (UC), that is root mean frequency differences between PM and UPM frequencies.}
    \label{fig:dev_plot}
\end{figure*}

\subsection{Variation of the coefficients in the $\Delta\nu - T_{\rm eff}-\log g-\kappa$ space}
\label{subsection:evol_fit_params}

\subsubsection{Prescriptions for radial modes}
\label{subsubsection:radial_modes}

In order to fit the parameters of the correction functions and to determine the fitting parameters, we used a least square minimization algorithm implementing a Levenberg-Marquardt method which minimizes the squared deviation defined as
\begin{equation}
   \mathcal{D} = \sum_{i = 1}^N\left[\frac{\nu_{{\rm PM, }i} - \nu_{{\rm UPM, }i} - \delta \nu_i}{\nu_{\rm max}}\right]^2 \, , 
\end{equation}
where $i$ corresponds to the eigenmode index, $N$ to the total number of radial modes and $\delta \nu$ the correction computed from the considered correction relation. Tables \ref{table:coeff_fit} and \ref{table:devscoeff_fit} summarize the coefficients and their squared deviations from our computations. We also define $\sqrt{\mathcal{D} / N}$ as the root-mean frequency differences after correction.

Coefficients $a$ and $b$ involved in Eq. \eqref{eq:kjeldsen} and $a_{3\rm , BG2}$ from Eq. \eqref{eq:combined} are presented in $T_{\rm eff} - \log g$ plane in Fig. \ref{fig:a_b_b4}. Furthermore, all coefficients are represented as a function of $\log z_1$ in Fig. \ref{fig:coeff_fit}. $a$, $\alpha$ and $a_{3\rm , BG2}$ (and therefore $a_{3\rm , BG1}$) show similar trends. Indeed, they are related to the amplitude of the surface effect at $\nu_{\rm max}$ and Eq. \eqref{eq:z1} allows one to understand their variations. However, this theoretical justification for the variations of $a$, $\alpha$, $a_{3\rm , BG1}$ and $a_{3\rm , BG2}$ does not provide a way to favour one correction law over an other. Coefficient $a_{-1\rm, BG2}$ exhibits the same behaviour as above. However, we cannot offer the same explanation for its trend because the inverse term in BG2 is a second correction to the cubic term and is not related to the amplitude of the surface effect at $\nu_{\rm max}$.

The trends followed by $b$ and $\beta$ are related to the slope of the frequency differences. As shown in \citet{Sonoi2015} and in Fig. \ref{fig:a_b_b4}, the coefficients $b$ (whatever the metallicity) increase significantly towards cooler stars, which again contradicts the assumption of a constant $b$. Regarding the relevance of giving a prescription for $\log b$ and $\log \beta$ thanks to the linear relationship with $\log z_1$, we can see in Fig. \ref{fig:coeff_fit} that $\log b$ and $\log \beta$ are affected by a high dispersion compared to the grey line. This could mean either that we omitted a physical dependency in Eq. \eqref{eq:z1} that only affects the agreement with $\log b$ and $\log \beta$, or a prescription based on same other physical basis should be investigated.

Table \ref{table:prescriptions} shows the prescriptions for the variations in the $T_{\rm eff}-\log g-\kappa$ space of all coefficients $c_0$ studied in this article in the form
\begin{align}
         \log c_0 =~& c_1 \log \Delta\nu/\Delta\nu_\odot + c_2 \log T_{\rm eff}/T_{\rm eff\odot} \nonumber \\ & + c_3 \log g/g_\odot + c_4 \log \kappa/\kappa_\odot +c_5
\end{align} 

We note that the opacity has a strong impact on each of the coefficients and must, therefore be taken into account when correcting the surface effect.

The top panel of Fig. \ref{fig:dev_plot}, top panel also shows the value of the root-mean frequency differences after correction for each model and each correction law. From this, we see that BG1 is the worst performer followed by K08. Those laws provide a correction that leaves frequency residuals comprised between 1 and $10~\mu\rm Hz$ which are still higher than the frequency resolution provided by CoRoT and \textit{Kepler}. The better performance of K08 over BG1 can be explained by the fact that K08 have two degrees of freedom whereas BG1 has only one. For radial modes, the inclusion of the normalized mode mass $E_{n\ell}$ in BG1 does not compensate the loss of a degree of freedom.

Then, the remaining laws K08r, S15 and BG2 provide correction almost as good as the resolution of CoRoT and \textit{Kepler}. K08r and BG2 are slightly better than S15, yet K08r is applied only \emph{on the frequency range} $0 < \nu/\nu_{\rm max} < 1.05$.

\subsubsection{Mixed-modes case}
\label{subsubsection:mixed_modes}

We also performed the same test as in \citet{Ball2017} on evolved models that present mixed-modes in their frequency spectrum. In subsection \ref{subsubsection:ball} we see that, thanks to the dependence in the normalized mode inertia, BG1 and BG2 can be applied to non-radial modes without any change of the law. However, in order to be able to compare all empirical corrections on non-radial modes, one had to rescale the frequency differences on which K08, K08r and S15 by mean of the inertia ratio $Q_{nl}$ for a mode of frequency $\nu_{nl}$ defined as the ratio of the inertia of this mode by the inertia of a radial mode interpolated at the frequency $\nu_{n\ell}$: $Q_{n\ell} = E_{n\ell} / E_{n0}(\nu_{n\ell})$ \citep[eg.][]{Rosenthal1999}. Furthermore, we added one last empirical relation by modifying the expression given for S15 in Eq. \eqref{eq:lorentz} similarly to BG1 and BG2 in which we replaced $\alpha$ by $\alpha/ E$ where $E$ is defined in Eq:\eqref{eq:mode_mass} (the new function is denoted S15E). This allows S15E to be applied directly on non-radial modes frequency differences.

The empirical relations K08, K08r and S15 were then adjusted on $Q_{n\ell}\delta \nu_{n\ell}$, with $0 \leq \ell \leq 2$ and S15E, BG1 and BG2 were adjusted directly on $\delta \nu_{n\ell}$, with $0 \leq \ell \leq 2$. For 9 of the 16 evolved models considered, the least-square algorithm converge to a solution of K08, K08r or S15 very remote from the general trend of frequency differences, whereas for the second group of relations (S15E, BG1 and BG2), the residual root-mean frequency differences after correction are greatly improved to a value between $0.1 - 1 ~\mu\mbox{Hz}$. 

As a third test, we performed the same fits excluding the quadrupolar modes (i.e. we fit modes with $0 \leq \ell \leq 1$). This time, corrections laws accuracies are similar to the one presented in section \ref{subsubsection:radial_modes}. As for the newly introduced S15E, it performs slightly worse than S15 but still better than K08. This third test suggests that the failure of K08, K08r and S15 in fitting $Q_{n\ell}\delta \nu_{n\ell}$, with $0 \leq \ell \leq 2$ is due to quadripolar modes. There are two reasons for it. 

\begin{figure}
    \centering
    \includegraphics[width=\hsize]{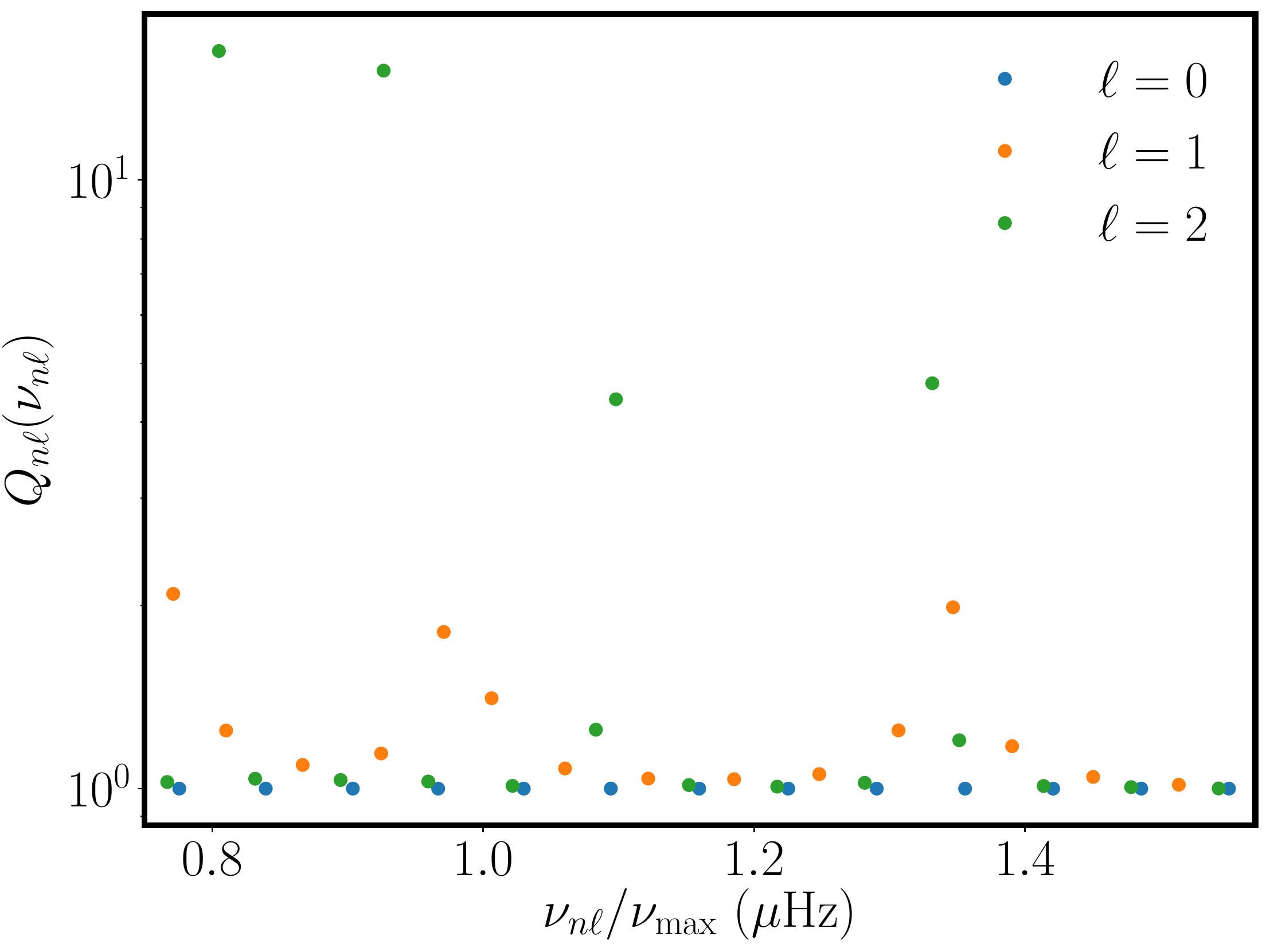}
    \caption{Ratio $Q_{n\ell}$ of a mode of degree $n\ell$ against frequency of the same mode normalized by $\nu_{\rm max}$ (only $\nu_{n\ell}/\nu_{\rm max} > 0.75$ for clarity), for the sub-giant model Cm10 with $T_{\rm eff} = 6503 ~\mbox{K}$, $\log g = 4.0$ and $[\rm Fe/H] = -1.0$. Each colour corresponds to a degree $\ell$. Yellow (resp. green) dots breaking from the general trend correspond to dipolar (resp. quadrupolar) mixed modes.}
    \label{fig:ekin}
\end{figure}

First, the p and g cavity are less coupled for $\ell = 2$ than for $\ell = 1$ mixed-modes which induces more important changes on the behaviour of a mode when the surface layers are changed between UPM and PM. Indeed, modifying the surface layers changes the frequency of pure p modes that couple with different g modes for PM and UPM \citep{Ball2017}. As a consequence, when computing $Q_{n,2}\delta \nu_{n,2}$, it so happens that we deal with mixed-modes from PM and UPM that have  different properties. Second, due to the presence of mixed-modes, $Q_{n, 2}$ sometimes becomes higher than ten, while it is normally of the order of unity (see example of Cm10 in Fig. \ref{fig:ekin}). It over-scales the corresponding quadrupolar mixed-modes and gives much weight to those modes, which in turns has strong impact on the quality of the fit. The peaks in the value of $Q_{n, 2}$ arise for mixed-modes having most of their amplitude in the g mode cavity, contrary to dipolar mixed-modes which have their $Q_{n, 1}$ staying close to unity. On the other hand, fitting directly $\delta \nu_{n\ell}$ with S15E, BG1 and BG2 does not amplify the frequency differences affecting mixed-modes, providing a much better correction.\\

For all these reasons, we recommend the use of BG2 or S15E when correcting sets of radial and non-radial modes and we recommend either S15 or BG2 when correcting only radial modes (K08r can be used for frequencies $\lesssim \nu_{\rm max}$). The remaining advantages of S15 over BG2 is that the coefficients $\alpha$ and $\beta$ are better described as a function of Eq. \eqref{eq:z1} than coefficients of BG2 which present a bigger dispersion on Fig. \ref{fig:coeff_fit}. This being said, many of the correction laws considered in this paper gives a root-mean frequency difference of the order of $0.1 ~\mu\mbox{Hz}$ (at least for few models), similar to the frequency uncertainties of CoRoT and \textit{Kepler}. Furthermore, the search for the best a posteriori correction law should not set aside the need of a theoretical understanding of the surface effect.

\section{Conclusion}
\label{section:s5}
We have computed a grid of 29 couples of one dimensional models using the method of patched models consisting in replacing poorly modelled surface layers of a 1D model by the stratification, averaged over the geometrical depth and time, computed from 3D hydrodynamical models. 
The grid includes models with effective temperature ranging from $T_{\rm eff} = 5\,000$ K to $6\,800$ K, surface gravity ranging from $\log g = 3.5$ to $4.5$ and iron abundance ranging from $[\rm Fe/H] = -1.0$ to $+0.5$.

Our aim was to estimate and understand the impact of varying metallicities on the surface effect. Our main result is that, in the considered range of metallicities (i.e. $[\rm Fe/H] = -1.0$ to $+0.5$) the amplitude of the surface effect computed at $\nu_{\rm max}$, and for models with same effective temperature and same surface gravity, can be up to a factor of three between the model with the lowest amplitude and the model with the highest one. However, it appears that studying the amplitude as a function of the metallicity does not lead to a clear trend, whereas the Rosseland mean opacity $\kappa$ turned out to be the adapted quantity for understanding the variation of the surface effect. Based on relatively simple physical arguments, consolidated using the grid of 3D models, we found a scaling relation between the amplitude of the surface effect and the global parameters $T_{\rm eff}$, $\log g$ and the opacity $\kappa$ computed at the photosphere. 

We also tested the accuracy of existing surface effect empirical corrections of radial modes frequency differences on each model of our grid in order to obtain a prescription for the coefficients. Then, we tested those laws on radial and non-radial modes for evolved models exhibiting mixed-modes, in order to test how the empirical corrections perform when mixed-modes are involved. Overall, the combined correction law proposed by \citet{Ball2014} is found to give the best performer, closely followed by the law proposed by \cite{Sonoi2015}. These two laws leave frequency differences that are less than $1~\mu \rm Hz$ on average, even reaching $0.1~\mu \rm Hz$ for the coolest stars of our set of model, which is of the order the frequency resolution provided by CoRoT and \textit{Kepler}. We note that, on a low frequency range ($0 < \nu/\nu_{\rm max} < 1.05$), the \citet{Kjeldsen2008} power law (calibrated on this reduced range) gives equivalent results. Then the \citet{Kjeldsen2008} power law calibrated on the whole range of frequency and the purely cubic correction proposed by \citet{Ball2014} are the worst performer with remaining mean frequency differences of the order of few $\mu \rm Hz$. When applying those corrections on frequency spectra including mixed-modes, only the empirical corrections BG1 and BG2 proposed by \citet{Ball2014} and the modified S15E where we added a factor of $1/E$ improve the mean frequency dispersion. Only S15E and BG2 leave a satisfying root-mean dispersion of the order of the CoRoT and \textit{Kepler} frequency resolution.

Therefore, we derived prescriptions for the fitting parameters of those radial modes correction empirical models as functions of $\log \Delta\nu$, $\log T_{\rm eff}$, $\log g$ and $\log \kappa$ which are quantities easily computed by 1D stellar evolution model. The next step will be to test our prescriptions against observed frequency spectrum in order to determine their degree of accuracy. We will focus on this in a furture work. 

Finally, we only considered in this article the issue of structural effects. However, other effects such as non-adiabaticity effects may also play a non-negligible role in the propagation of acoustic waves in the surface layers. This will be studied in a forthcoming paper. 

\begin{acknowledgements}
    L.M. thanks Dr C. Pinçon for many interesting discussions.
    H.G.L. acknowledges financial support by the Sonderforschungsbereich SFB\,881 ``The Milky Way System'' (sub-projects A4) of the German Research Foundation (DFG).
\end{acknowledgements}

    \bibliographystyle{aa} 
    \bibliography{biblio}

    \begin{appendix}
        \section{Derivation of the scaling relation (Eq. \eqref{eq:deltanu_elevation})}
\label{appendix:app1}

We start from Eq. \eqref{eq:deltanu_kernel}. After \citet{Christensen-Dalsgaard1997}, we have 
\begin{eqnarray}
        \tilde{K}_{c^2, v}^{nl} & \simeq & 0 \nonumber\\
        \tilde{K}_{v, c^2}^{nl} &=& \omega^{-2}\displaystyle{\left(\int_0^M \xi^2_r {\rm d}m \right)^{-1} \left(\frac{{\rm d}\xi_r }{{\rm d}r }\right)^2 4\pi r^2 \rho c_{\rm s}^2},
\end{eqnarray}
where $\omega = 2\pi \nu$ and $\xi_r$ are respectively the angular frequency and the radial displacement of the mode, $c_{\rm s}$ is the sound speed and $M$ and $R$ are respectively the total mass and the total radius of the star.

The first term in parentheses in the expression of $\tilde{K}_{v, c^2}^{nl}$ is the inertia of a mode: 
\begin{equation}
    \mathcal{I} \simeq \int_0^M\xi_r^2 {\rm d}m = \int_0^R\xi_r^2 4\pi r^2 \rho{\rm d}r.
    \label{eq:inertia}
\end{equation}

Using a first-order expansion, we can write the radial displacement as (e.g. \citet{Unno1989}):
\begin{equation}
\xi_r(r) \simeq A \rho^{-\nicefrac{1}{2}}c_{\rm s}^{-\nicefrac{1}{2}}r^{-1}\cos\left( \omega\int_0^r\frac{{\rm d }r'}{c_{\rm s}} - \zeta \right),
\label{eq:rad_disp}
\end{equation}
where $A$ is a constant and $\zeta$ is a phase factor. Further inserting Eq. \eqref{eq:rad_disp} into Eq. \eqref{eq:inertia} leads to
\begin{equation}
    \mathcal{I} = \int_0^R 4\pi A^2 \cos^2\left( \omega\int_0^r\frac{{\rm d }r'}{c_{\rm s}} - \zeta \right)  \frac{{\rm d}r}{c_{\rm s}},
\end{equation}
Averaging the cosine term gives $\nicefrac{1}{2}$ and then simply
\begin{equation}
    \mathcal{I} = 2\pi A^2\int_0^R \frac{{\rm d}r}{c_{\rm s}} = \frac{\pi A^2}{\Delta \nu},
    \label{eq:inertia_simple}
\end{equation}
with $\Delta \nu$ defined by $\Delta \nu = \left(2\int_0^R {{\rm d}r}/{c_{\rm s}}\right)^{-1}$.
Then, $\tilde{K}_{v, c^2}^{nl}$ reads
\begin{equation}
    \tilde{K}_{v, c^2}^{nl} = \frac{\omega^{-2}\Delta \nu}{\pi A^2}\left(\frac{{\rm d}\xi_r }{{\rm d}r }\right)^2 4\pi r^2 \rho c_{\rm s}^2.
\end{equation}
Yet, $\left({{\rm d}\xi_r }/{{\rm d}r }\right)^2 \simeq k_r^2 \xi_r^2 = \omega^2 \xi_r^2 / c_{\rm s}^2$. Then,
\begin{equation}
    \begin{array}{lll}
        \tilde{K}_{v, c^2}^{nl} &=& \displaystyle{\frac{\omega^{-2}\Delta \nu}{\pi A^2} \frac{\omega^2}{c_{\rm s}^2}} 4\pi r^2 \rho c_{\rm s}^2\xi_r^2 \\
                                &=& \displaystyle{\frac{4\Delta \nu}{A^2}}r^2\rho A^2\rho^{-1}c_{\rm s}^{-1}r^{-2}\cos^2\left( \omega\int_0^r\frac{{\rm d }r'}{c_{\rm s}} - \zeta\right),
    \end{array}
\end{equation}
where the last line was obtained by replacing $\xi_r^2$ by its expression. Finally, by simplifying this expression we obtain:
\begin{equation}
    \tilde{K}_{v, c^2}^{nl} \simeq \frac{2 \Delta \nu}{c_{\rm s}}\cos^2\left( \omega\int_0^r\frac{{\rm d }r'}{c_{\rm s}} - \zeta\right).
    \label{eq:approx_K}
\end{equation}
Eventually, inserting Eq. \eqref{eq:approx_K} into Eq. \eqref{eq:deltanu_kernel}, approximating the cosine by $\nicefrac{1}{2}$ as in \eqref{eq:inertia_simple} and using Eq. (19) from \citet{Rosenthal1999}: 
\begin{equation}
    \int_0^R\frac{\delta_m v}{v} \frac{{\rm d} r}{c_{\rm s}} \simeq \frac{ \Delta r}{2c_{\rm s, ph}},
\end{equation}
with $c_{\rm s, ph}$ the sound speed at the photosphere, we obtain the expression of the amplitude of the surface effect proposed in \citet{Rosenthal1999}
\begin{equation}
     \frac{\delta \nu}{\nu} \simeq \frac{\Delta \nu \Delta r}{c_{\rm s, ph}}.
     \label{eq:deltanu_rosenthal_final}
\end{equation}

\section{Fitting coefficients}
\label{appendix:2}

In Table \ref{table:coeff_fit} we gather the values of the fitting parameters introduced in Eqs. \eqref{eq:kjeldsen}, \eqref{eq:cubic}, \eqref{eq:combined}, and \eqref{eq:lorentz} used in order to perform the fits shown in Fig. \ref{fig:coeff_fit} and toderive the coefficients for the prescriptions given in Table \ref{table:prescriptions}.

\begin{table*}
    \caption{Fitting parameters of K08, K08r, S15, BG1, and BG2.}  
    \label{table:coeff_fit}     
    \centering 
{\renewcommand{\arraystretch}{1.15}
\begin{tabular}{l | c c | c c | c c | c | c c}       
    \hline\hline                   
    & \multicolumn{2}{c|}{K08} & \multicolumn{2}{c|}{K08r} & \multicolumn{2}{c|}{S15} & \multicolumn{1}{c|}{BG1} & \multicolumn{2}{c}{BG2}\\
    Model & $|a|$ &   $b$ & $|a|$ &   $b$ & $|\alpha / 2|$ & $\beta$ & $|a_{3\rm , BG1}|$ & $|a_{-1\rm , BG2}|$ & $|a_{3\rm , BG2}|$\\
    \hline
    Cp05 & $9.19 \times 10^{-3}$ & $1.55$ & $1.06 \times 10^{-2}$ & $2.17$ & $1.01 \times 10^{-2}$ & $2.98$ & $-9.66 \times 10^{-12}$ &$-7.99 \times 10^{-12}$ &$-6.01 \times 10^{-12}$\\
    Fp05 & $7.13 \times 10^{-3}$ & $1.43$ & $9.66 \times 10^{-3}$ & $2.74$ & $7.90 \times 10^{-3}$ & $3.03$ & $-8.03 \times 10^{-12}$ &$-7.98 \times 10^{-12}$ &$-5.17 \times 10^{-12}$\\
    Jp05 & $5.64 \times 10^{-3}$ & $1.64$ & $7.30 \times 10^{-3}$ & $2.95$ & $6.32 \times 10^{-3}$ & $3.56$ & $-6.94 \times 10^{-12}$ &$-7.23 \times 10^{-12}$ &$-4.31 \times 10^{-12}$\\
    Lp05 & $4.79 \times 10^{-3}$ & $1.5$  & $6.03 \times 10^{-3}$ & $2.75$ & $5.33 \times 10^{-3}$ & $3.36$ & $-6.03 \times 10^{-12}$ &$-8.41 \times 10^{-12}$ &$-3.38 \times 10^{-12}$\\
    Op05 & $1.63 \times 10^{-3}$ & $1.74$ & $1.98 \times 10^{-3}$ & $4$    & $1.80 \times 10^{-3}$ & $5.23$ & $-3.30 \times 10^{-12}$ &$-4.03 \times 10^{-12}$ &$-2.08 \times 10^{-12}$\\
    \hline
    Am00 & $2.05 \times 10^{-3}$ & $1.82$ & $2.39 \times 10^{-3}$ & $4.1$  & $2.27 \times 10^{-3}$ & $6$    & $-1.24 \times 10^{-12}$ &$-1.01 \times 10^{-12}$ &$-9.19 \times 10^{-13}$\\
    Bm00 & $9.36 \times 10^{-3}$ & $1.35$ & $1.14 \times 10^{-2}$ & $2.12$ & $1.02 \times 10^{-2}$ & $2.65$ & $-4.66 \times 10^{-12}$ &$-4.15 \times 10^{-12}$ &$-2.96 \times 10^{-12}$\\
    Cm00 & $8.93 \times 10^{-3}$ & $1.49$ & $1.12 \times 10^{-2}$ & $2.51$ & $9.96 \times 10^{-3}$ & $3.09$ & $-1.12 \times 10^{-11}$ &$-8.32 \times 10^{-12}$ &$-7.99 \times 10^{-12}$\\
    Dm00 & $6.66 \times 10^{-3}$ & $1.43$ & $8.58 \times 10^{-3}$ & $2.65$ & $7.40 \times 10^{-3}$ & $3.16$ & $-4.19 \times 10^{-12}$ &$-3.92 \times 10^{-12}$ &$-2.84 \times 10^{-12}$\\
    Fm00 & $7.26 \times 10^{-3}$ & $1.5$  & $9.54 \times 10^{-3}$ & $2.82$ & $8.08 \times 10^{-3}$ & $3.29$ & $-1.03 \times 10^{-11}$ &$-9.16 \times 10^{-12}$ &$-7.00 \times 10^{-12}$\\
    Gm00 & $1.41 \times 10^{-2}$ & $1.77$ & $1.53 \times 10^{-2}$ & $2$    & $1.47 \times 10^{-2}$ & $2.85$ & $-9.66 \times 10^{-11}$ &$-6.73 \times 10^{-11}$ &$-4.44 \times 10^{-11}$\\
    Hm00 & $4.61 \times 10^{-3}$ & $1.52$ & $5.76 \times 10^{-3}$ & $2.93$ & $5.11 \times 10^{-3}$ & $3.59$ & $-3.63 \times 10^{-12}$ &$-3.53 \times 10^{-12}$ &$-2.42 \times 10^{-12}$\\
    Im00 & $1.62 \times 10^{-3}$ & $2.02$ & $1.90 \times 10^{-3}$ & $5.5$  & $1.78 \times 10^{-3}$ & $8.09$ & $-8.51 \times 10^{-13}$ &$-6.08 \times 10^{-13}$ &$-6.50 \times 10^{-13}$\\
    Jm00 & $3.06 \times 10^{-3}$ & $1.88$ & $3.88 \times 10^{-3}$ & $4.54$ & $3.40 \times 10^{-3}$ & $5.75$ & $-5.79 \times 10^{-12}$ &$-4.12 \times 10^{-12}$ &$-4.30 \times 10^{-12}$\\
    Km00 & $9.69 \times 10^{-3}$ & $1.5$  & $1.32 \times 10^{-2}$ & $2.79$ & $1.08 \times 10^{-2}$ & $3.13$ & $-7.23 \times 10^{-11}$ &$-6.07 \times 10^{-11}$ &$-4.88 \times 10^{-11}$\\
    Mm00 & $6.61 \times 10^{-3}$ & $1.53$ & $8.45 \times 10^{-3}$ & $2.8$  & $7.35 \times 10^{-3}$ & $3.36$ & $-6.35 \times 10^{-11}$ &$-6.43 \times 10^{-11}$ &$-4.03 \times 10^{-11}$\\
    \hline
    Em05 & $2.41 \times 10^{-3}$ & $1.7$  & $2.92 \times 10^{-3}$ & $3.43$ & $2.68 \times 10^{-3}$ & $4.38$ & $-1.09 \times 10^{-12}$ &$-1.20 \times 10^{-12}$ &$-6.75 \times 10^{-13}$\\
    Fm05 & $7.00 \times 10^{-3}$ & $1.42$ & $9.11 \times 10^{-3}$ & $2.59$ & $7.71 \times 10^{-3}$ & $3$    & $-1.02 \times 10^{-11}$ &$-1.16 \times 10^{-11}$ &$-5.92 \times 10^{-12}$\\
    Gm05 & $1.03 \times 10^{-2}$ & $1.51$ & $1.34 \times 10^{-2}$ & $2.47$ & $1.12 \times 10^{-2}$ & $2.83$ & $-7.51 \times 10^{-11}$ &$-6.21 \times 10^{-11}$ &$-4.34 \times 10^{-11}$\\
    Im05 & $1.60 \times 10^{-3}$ & $2.45$ & $1.79 \times 10^{-3}$ & $5.29$ & $1.97 \times 10^{-3}$ & $7.72$ & $-1.15 \times 10^{-12}$ &$-3.76 \times 10^{-13}$ &$-1.02 \times 10^{-12}$\\
    Jm05 & $3.44 \times 10^{-3}$ & $1.67$ & $4.25 \times 10^{-3}$ & $3.52$ & $3.82 \times 10^{-3}$ & $4.38$ & $-7.33 \times 10^{-12}$ &$-6.72 \times 10^{-12}$ &$-4.98 \times 10^{-12}$\\
    Km05 & $1.16 \times 10^{-2}$ & $1.38$ & $1.41 \times 10^{-2}$ & $2.15$ & $1.27 \times 10^{-2}$ & $2.73$ & $-1.03 \times 10^{-10}$ &$-1.15 \times 10^{-10}$ &$-5.85 \times 10^{-11}$\\
    Mm05 & $4.55 \times 10^{-3}$ & $1.7$  & $5.76 \times 10^{-3}$ & $3.38$ & $5.09 \times 10^{-3}$ & $4.1$  & $-5.81 \times 10^{-11}$ &$-5.01 \times 10^{-11}$ &$-3.97 \times 10^{-11}$\\
    \hline
    Bm10 & $6.10 \times 10^{-3}$ & $1.37$ & $7.93 \times 10^{-3}$ & $2.48$ & $6.64 \times 10^{-3}$ & $2.77$ & $-3.21 \times 10^{-12}$ &$-3.59 \times 10^{-12}$ &$-1.78 \times 10^{-12}$\\
    Am10 & $2.45 \times 10^{-3}$ & $2.49$ & $2.56 \times 10^{-3}$ & $4.79$ & $3.04 \times 10^{-3}$ & $7.68$ & $-4.00 \times 10^{-12}$ &$-7.97 \times 10^{-13}$ &$-3.72 \times 10^{-12}$\\
    Cm10 & $8.64 \times 10^{-3}$ & $1.41$ & $1.13 \times 10^{-2}$ & $2.57$ & $9.51 \times 10^{-3}$ & $2.95$ & $-1.11 \times 10^{-11}$ &$-1.11 \times 10^{-11}$ &$-6.83 \times 10^{-12}$\\
    Fm10 & $5.89 \times 10^{-3}$ & $1.54$ & $7.47 \times 10^{-3}$ & $2.96$ & $6.55 \times 10^{-3}$ & $3.57$ & $-1.24 \times 10^{-11}$ &$-1.11 \times 10^{-11}$ &$-8.41 \times 10^{-12}$\\
    Gm10 & $1.48 \times 10^{-2}$ & $1.36$ & $1.89 \times 10^{-2}$ & $2.3$  & $1.64 \times 10^{-2}$ & $2.76$ & $-1.04 \times 10^{-10}$ &$-7.87 \times 10^{-11}$ &$-7.35 \times 10^{-11}$\\
    Km10 & $1.04 \times 10^{-2}$ & $1.58$ & $1.23 \times 10^{-2}$ & $2.54$ & $1.15 \times 10^{-2}$ & $3.4$  & $-1.11 \times 10^{-10}$ &$-8.96 \times 10^{-11}$ &$-7.59 \times 10^{-11}$\\
    \hline
\end{tabular}}
\tablefoot{Coefficients $a$, $\alpha$, $a_{3\rm ,BG1}$, $a_{-1\rm , BG2}$ and $a_{3\rm , BG2}$ are all negative.}
\end{table*}

\section{Root-mean frequency differences after correction}
\label{appendix:3}

In Table \ref{table:devscoeff_fit} we gather the values of the root-mean frequency differences $\nu_{\rm max}\sqrt{\mathcal{D}/N}$ of radial modes after correction for each empirical law shown in Fig. \ref{fig:dev_plot}.

\begin{table*}
    \caption{Root-mean frequency differences $\nu_{\rm max}\sqrt{\mathcal{D} / N} ~[\rm \mu Hz]$ of radial modes after correction for each empirical law K08, K08r, S15, BG1, and BG2 and for each model. UC denotes $\nu_{\rm max}\sqrt{\mathcal{D} / N}$ for uncorrected frequencies.}
    \label{table:devscoeff_fit}     
    \centering                         
    \begin{tabular}{l | c c c c c c}       
        \hline\hline                   
                & UC     & K08       & K08r          & S15           & BG1           & BG2\\
        \hline
        Cp05    & $9.11$ & $1.4$     & $1.03$        & $0.808$       & $2.83$        & $0.398$ \\
        Fp05    & $7.99$ & $1.28$    & $0.443$       & $0.675$       & $2.39$        & $0.26$ \\
        Jp05    & $6.5$  & $0.931$   & $0.347$       & $0.365$       & $1.87$        & $0.279$ \\
        Lp05    & $5.98$ & $0.917$   & $0.311$       & $0.369$       & $1.95$        & $0.308$ \\
        Op05    & $2.3$  & $0.456$   & $0.0371$      & $0.175$       & $0.635$       & $0.158$ \\
        \hline 
        Am00    & $7.49$ & $1.63$    & $0.249$       & $0.726$       & $1.45$        & $0.272$ \\
        Bm00    & $16.8$ & $2.27$    & $1.06$        & $1.23$        & $5.15$        & $1.36$ \\
        Cm00    & $9.85$ & $1.42$    & $0.692$       & $0.806$       & $2.56$        & $0.719$ \\
        Dm00    & $13.4$ & $2$       & $0.709$       & $0.848$       & $3.66$        & $1.02$ \\
        Fm00    & $8.29$ & $1.3$     & $0.455$       & $0.562$       & $2.22$        & $0.474$ \\
        Gm00    & $3.47$ & $0.437$   & $0.432$       & $0.306$       & $1.26$        & $0.163$ \\
        Hm00    & $9.6$  & $1.57$    & $0.283$       & $0.667$       & $2.48$        & $0.553$ \\
        Im00    & $6.82$ & $1.49$    & $0.161$       & $0.419$       & $1.28$        & $0.567$ \\
        Jm00    & $3.78$ & $0.822$   & $0.195$       & $0.334$       & $0.8$         & $0.313$ \\
        Km00    & $3.5$  & $0.534$   & $0.21$        & $0.267$       & $0.974$       & $0.112$ \\
        Mm00    & $2.53$ & $0.39$    & $0.14$        & $0.163$       & $0.728$       & $0.1$ \\
        \hline 
        Em05    & $8.88$ & $1.69$    & $0.186$       & $0.761$       & $2.36$        & $0.337$ \\
        Fm05    & $7.7$  & $1.29$    & $0.517$       & $0.657$       & $2.48$        & $0.475$ \\
        Gm05    & $3.17$ & $0.504$   & $0.205$       & $0.305$       & $1.02$        & $0.146$ \\
        Im05    & $7.48$ & $1.05$    & $0.181$       & $0.551$       & $0.691$       & $0.361$ \\
        Jm05    & $4.18$ & $0.805$   & $0.0858$      & $0.36$        & $1.01$        & $0.17$ \\
        Km05    & $4.19$ & $0.62$    & $0.392$       & $0.329$       & $1.42$        & $0.388$ \\
        Mm05    & $1.78$ & $0.301$   & $0.0532$      & $0.117$       & $0.432$       & $0.0624$ \\
        \hline 
        Bm10    & $10.7$ & $1.68$    & $0.428$       & $1.03$        & $3.6$         & $0.791$ \\
        Am10    & $10.1$ & $1.38$    & $0.219$       & $0.89$        & $0.558$       & $0.229$ \\
        Cm10    & $9.3$  & $1.55$    & $0.59$        & $0.838$       & $2.85$        & $0.454$ \\
        Fm10    & $6.77$ & $1.16$    & $0.333$       & $0.511$       & $1.72$        & $0.308$ \\
        Gm10    & $5.29$ & $0.727$   & $0.32$        & $0.388$       & $1.46$        & $0.478$ \\
        Km10    & $3.82$ & $0.544$   & $0.245$       & $0.2$         & $0.975$       & $0.193$ \\
        \hline
        \end{tabular}
\end{table*}

    \end{appendix}

\end{document}